\documentclass[aps,preprintnumbers,showpacs,nofootinbib,superscriptaddress,floatfix]{revtex4}

\pdfoutput=1 

\usepackage{graphicx}
\usepackage{amssymb}
\usepackage{amsmath}
\usepackage{bm}
\usepackage{slashed}
\usepackage{datetime}
\usepackage{mciteplus}
\usepackage{multirow}
\usepackage{color}
\usepackage[utf8]{inputenc}
\usepackage[normalem]{ulem}

\newcommand{\smarrow}{\mbox{\raisebox{-4.5pt}[0pt][0pt]{$\hspace{-1pt} 
		\vec{\phantom{v}}$}}}

\newcommand{\hermes}{\textsc{Hermes}}
\newcommand{\compass}{\textsc{Compass}}
\newcommand{\minuit}{\textsc{Minuit}}

\newcommand{\T}{\perp}
\newcommand{\Tperp}{T}
\newcommand{\bT}{\xi_T}
\newcommand{\bb}{\xi}

\begin{document}
\allowdisplaybreaks[2]

\title{
Extraction of partonic transverse momentum distributions 
from semi-inclusive 
deep-inelastic scattering, Drell--Yan and Z-boson production
}

\author{Alessandro Bacchetta}
\email{alessandro.bacchetta@unipv.it}
\affiliation{Dipartimento di Fisica, Universit\`a di Pavia, via Bassi 6,
  I-27100 Pavia} 
\affiliation{INFN Sezione di Pavia, via Bassi 6, I-27100 Pavia, Italy}

\author{Filippo Delcarro}
\email{filippo.delcarro@pv.infn.it}
\affiliation{Dipartimento di Fisica, Universit\`a di Pavia, via Bassi 6,
  I-27100 Pavia} 
\affiliation{INFN Sezione di Pavia, via Bassi 6, I-27100 Pavia, Italy}

\author{Cristian Pisano}
\email{cristian.pisano@unipv.it}
\affiliation{Dipartimento di Fisica, Universit\`a di Pavia, via Bassi 6,
  I-27100 Pavia} 
\affiliation{INFN Sezione di Pavia, via Bassi 6, I-27100 Pavia, Italy}

\author{Marco Radici}
\email{marco.radici@pv.infn.it}
\affiliation{INFN Sezione di Pavia, via Bassi 6, I-27100 Pavia, Italy}

\author{Andrea Signori}
\email{asignori@jlab.org}
\affiliation{Theory Center, Thomas Jefferson National Accelerator Facility, 12000 Jefferson Avenue, Newport News, VA 23606, USA}

\begin{abstract}
We present an extraction of unpolarized partonic transverse momentum
distributions (TMDs) 
from a simultaneous fit of available data measured in semi-inclusive 
deep-inelastic scattering, 
Drell--Yan and $Z$ boson production. 
To connect data at different scales, we use TMD evolution at next-to-leading logarithmic accuracy. The
analysis is restricted to the low-transverse-momentum region, with no matching
to fixed-order calculations at high transverse momentum. We introduce specific
choices to deal with TMD evolution at low scales, of the order of 1 GeV$^2$.
This could be considered as a first attempt at a global fit of TMDs.
\end{abstract}

\preprint{JLAB THY 17-2437}


\pacs{13.60.Le, 13.87.Fh,14.20.Dh}

\maketitle
\tableofcontents

\newpage
\section{Introduction}
\label{s:intro}

Parton distribution functions describe the internal structure of the nucleon
in terms of its elementary constituents (quarks and gluons). They cannot be
easily computed from first principles, because they require the ability to
carry out Quantum Chromodynamics (QCD) calculations in its nonperturbative
regime. Many experimental observables in hard scattering experiments
involving hadrons are related to parton distribution functions (PDFs) and
fragmentation functions (FFs), in a way that is specified by factorization
theorems (see, e.g., Refs.~\cite{Collins:1989gx,Collins:2011zzd}). 
These theorems also elucidate the universality properties of PDFs and FFs
(i.e., the fact that they are the same in different processes) 
and their evolution equations (i.e., how they get modified by the change in
the hard scale of the process). 
Availability of measurements of different processes in different
experiments makes it possible to test factorization
theorems and extract PDFs and FFs through so-called global fits. 
On the other side, the knowledge of PDFs and FFs allows us
to make predictions for other hard hadronic processes. 
These general statements apply equally well to
standard collinear PDFs and FFs and to transverse-momentum-dependent parton
distribution functions (TMD PDFs) and fragmentation functions (TMD FFs). 
Collinear PDFs
describe the distribution of partons integrated over all components of
partonic momentum except the one collinear to the parent hadron; hence,
collinear PDFs
are functions of the parton longitudinal momentum fraction $x$. 
TMD PDFs (or TMDs for short) 
include also the dependence on the transverse momentum $\bm{k}_{\T}$. 
They can be interpreted as three-dimensional generalizations of collinear PDFs.
Similar arguments apply to collinear FFs and TMD
FFs~\cite{Angeles-Martinez:2015sea}. 

There are several differences between collinear and TMD distributions. From
the formal point of view, factorization theorems for the two types of
functions are different, implying also different universality
properties and evolution equations~\cite{Rogers:2015sqa}. From the
experimental point 
of view, observables related to TMDs require the measurement of some transverse
momentum component much smaller than the hard scale of the
process~\cite{Bacchetta:2016ccz,Radici:2016hbh}.  For
instance, Deep-Inelastic Scattering (DIS) is characterized by a hard scale represented by the
4-momentum squared of the virtual photon ($-Q^2$). In inclusive DIS this is
the only scale of the process, and only collinear PDFs
and FFs can be accessed. In semi-inclusive DIS (SIDIS) also the transverse momentum of the
outgoing  
hadron ($P_{hT}$) can be measured~\cite{Mulders:1995dh,Bacchetta:2006tn}. 
If $P_{hT}^2\ll Q^2$, TMD
factorization can be applied and the process is sensitive to
TMDs~\cite{Collins:2011zzd}.

If polarization is taken into account, several TMDs can be
introduced~\cite{Mulders:1995dh,Boer:1997nt,
Bacchetta:2000jk,
Mulders:2000sh,
Boer:2016xqr
}. Attempts to extract some of them have already been presented in the past~\cite{Bacchetta:2011gx,Anselmino:2012aa,Echevarria:2014xaa,Anselmino:2016uie,
Lu:2009ip,Barone:2015ksa,
Lefky:2014eia,
Anselmino:2013vqa,Kang:2015msa
}.  In
this work, we focus on the simplest ones, i.e., the unpolarized TMD
PDF $f_1^q(x,k_{\T}^2)$ and the unpolarized TMD
FF $D_1^{q \to h}(z,P_{\perp}^2)$, where $z$ is
  the fractional energy carried by the detected hadron $h$, $k_{\T}$ is the
  transverse momentum of the parton with respect to the parent hadron, and
  $P_{\perp}$ is the transverse momentum of the produced hadron with
  respect to the parent parton. Despite their
  simplicity, the phenomenology of these unpolarized TMDs presents several
  challenges~\cite{Signori:2016lvd}: the choice of a functional form 
  for the nonperturbative components of TMDs, 
  the inclusion of a possible dependence on partonic
  flavor~\cite{Signori:2013mda}, the implementation of TMD
  evolution~\cite{Bacchetta:2015ora,Rogers:2015sqa}, the matching to
  fixed-order calculations in collinear
  factorization~\cite{Collins:2016hqq}. 

We take into consideration three kinds of processes: SIDIS, 
Drell--Yan processes (DY) and the production of $Z$
bosons. To date, they represent 
all possible processes
  where experimental information is available for unpolarized TMD
  extractions. 
The only important
process currently missing is electron-positron annihilation, which is
particularly important for the determination of TMD
FFs~\cite{Bacchetta:2015ora}. This work can therefore be considered as the
first attempt at a global fit of TMDs.  

The paper is organized as follows. In Sec.~\ref{s:theory}, the general
formalism for TMDs in SIDIS, DY processes, and Z production is briefly outlined, including a
description of the assumptions and approximations in the phenomenological
implementation of TMD evolution equations. In Sec.~\ref{s:data_analysis}, the criteria
for selecting the data analyzed in the fit are summarized and commented. In
Sec.~\ref{s:results}, the results of our global fit are presented and
discussed. In Sec.~\ref{s:conclusions}, we summarize the results and present an outlook for future improvements.

\section{Formalism}
\label{s:theory}

\subsection{Semi-inclusive DIS}
\label{ss:SIDIS_formalism}

In one-particle SIDIS, a lepton $\ell$ with momentum $l$ scatters 
off a hadron target $N$ with mass $M$ and momentum
$P$. In the final state, the scattered lepton momentum 
$l'$ is measured together with
one hadron $h$ with mass $M_h$
and momentum $P_h$. The corresponding reaction formula is  
\begin{equation}
  \label{e:sidis}
\ell(l) + N(P) \to \ell(l') + h(P_h) + X \, .
\end{equation}
The space-like momentum transfer is $q = l - l'$, with $Q^2 = - q^2$. We
introduce the usual invariants  
\begin{align}
  \label{e:xyz}
x &= \frac{Q^2}{2\,P\cdot q},
&
y &= \frac{P \cdot q}{P \cdot l},
&
z &= \frac{P \cdot P_h}{P\cdot q},
&
\gamma &= \frac{2 M x}{Q} .
\end{align}

The available data refer to SIDIS hadron multiplicities, namely to the differential number of hadrons produced per corresponding inclusive DIS event. In terms of cross sections, we define the multiplicities as
\begin{equation}
m_N^h (x,z,|\bm{P}_{h\Tperp}|, Q^2) = \frac{d \sigma_N^h / ( dx  dz d|\bm{P}_{h\Tperp}| dQ^2) }
                                                                   {d\sigma_{\text{DIS}} / ( dx dQ^2 ) }\, ,
\label{e:multiplicity}
\end{equation}
where $d\sigma_N^h$ is the differential cross section for the SIDIS process and $d\sigma_{\text{DIS}}$ is the corresponding inclusive one, 
and where \( \bm{P}_{h\Tperp} \) is the component of \( \bm{P}_{h} \)
transverse to \( \bm{q} \) (we follow here the notation suggested in
Ref.~\cite{Boer:2011fh}).  
In the single-photon-exchange approximation, the multiplicities can be written
as ratios of 
structure functions (see Ref.~\cite{Bacchetta:2006tn} for details):
\begin{equation}
m_N^h (x,z,|\bm{P}_{h\Tperp}|, Q^2) =   
\frac{2 \pi\,|\bm{P}_{h\Tperp}| F_{UU ,T}(x,z,\bm{P}_{h\Tperp}^2, Q^2) + 2 \pi
  \varepsilon |\bm{P}_{h\Tperp}| F_{UU ,L}(x,z,\bm{P}_{h\Tperp}^2, Q^2)}
        {F_{T}(x,Q^2) + \varepsilon  F_{L}(x,Q^2)} \, ,
 \label{e:mFF}
\end{equation} 
where
\begin{align}
\varepsilon &= \frac{1-y -\frac{1}{4} \gamma^2 y^2}{1-y+\frac{1}{2} y^2 +\frac{1}{4} \gamma^2 y^2} \ .
\end{align}  
In the numerator of Eq.~\eqref{e:mFF} the structure function $F_{XY,Z}$ corresponds to a lepton with polarization $X$ scattering on a target with polarization $Y$ by exchanging a virtual photon in a polarization state $Z$. 
In the denominator, only the photon polarization is explicitly written ($T$,
$L$), as usually done in the literature.

The semi-inclusive cross section can be expressed in a factorized form in
terms of TMDs only in the kinematic limits $M^2 \ll Q^2$ and 
$\bm{P}_{hT}^2 \ll Q^2$. 
In these limits, the structure function $F_{UU,L}$ of Eq.~\eqref{e:mFF}
can be neglected~\cite{Bacchetta:2008xw}. 
 The structure function $F_L$ in the denominator contains contributions
 involving powers of the strong coupling constant $\alpha_S$ at an order that
 goes beyond the level reached in this analysis; 
hence, it will be
 consistently neglected (for measurements and
 estimates of the $F_L$ structure function see, e.g.,
 Refs.~\cite{Chekanov:2009na,Andreev:2013vha} and references therein).  

\begin{figure}
\centering
\includegraphics[width=0.6\textwidth]{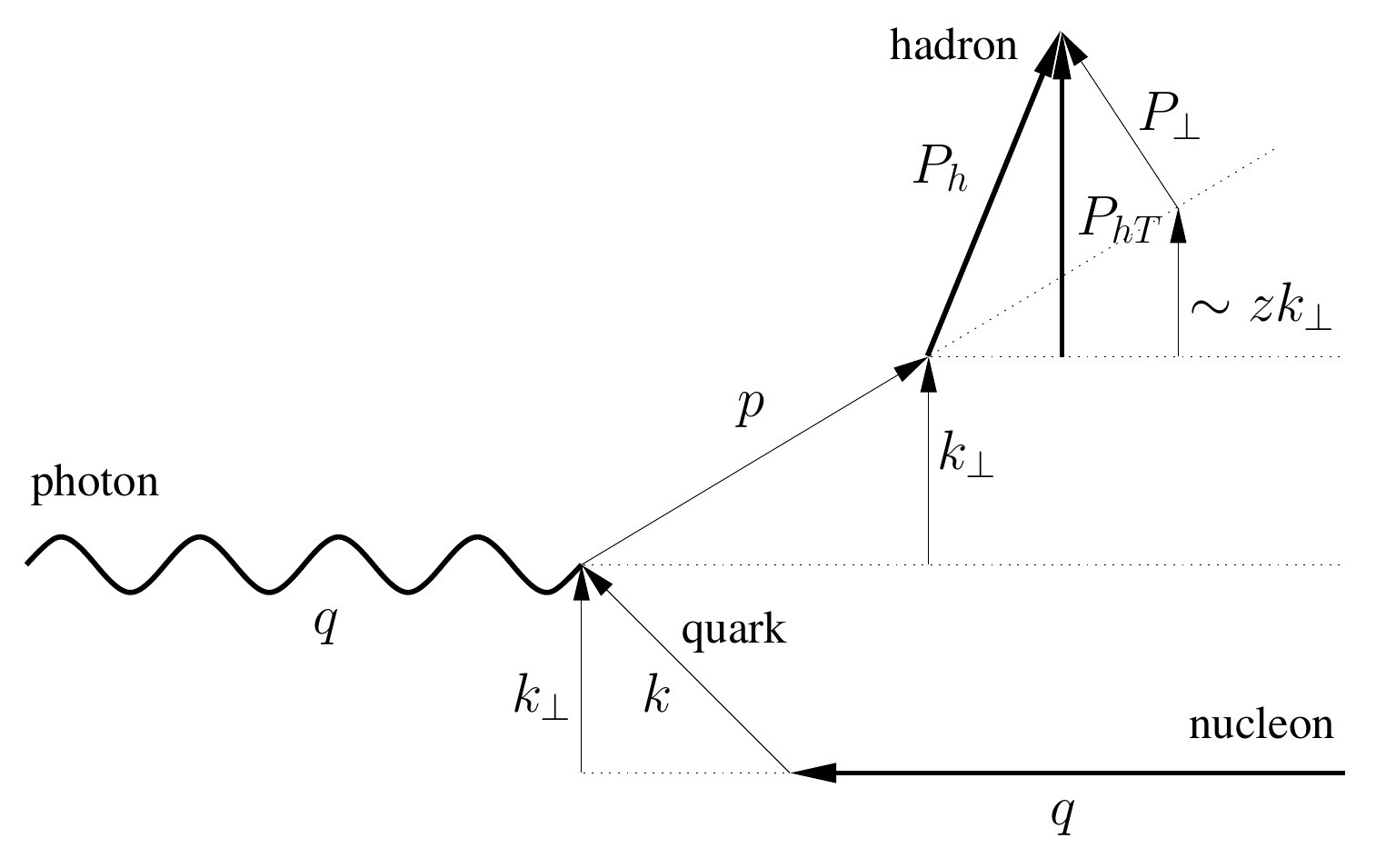}
\caption{Diagram describing the relevant momenta involved in a semi-inclusive DIS event (see also Ref.~\cite{Matevosyan:2011vj}): a
  virtual photon (defining the reference axis) strikes a parton inside a
  proton. The parton has a transverse momentum $\bm{k}_\perp$ (not measured). The
  struck parton fragments into a hadron, which acquires a further transverse
  momentum $\bm{P}_\perp$ (not measured). 
 The total measured transverse-momentum of the
  final hadron is $\bm{P}_{h\Tperp}$. When $Q^2$ is very large, the longitudinal
  components are all much larger than the transverse components. In this
  regime,  
  $\bm{P}_{h\Tperp} \approx z \bm{k}_\T + \bm{P}_\T$.} 
\label{f:trans_momenta_SIDIS}
\end{figure}

To express the structure functions in terms of TMD PDFs and FFs, 
we rely on the factorized formula 
for SIDIS~\cite{Collins:1981uk,Collins:1984kg,Ji:2002aa,Ji:2004wu,%
Collins:2011zzd,Aybat:2011zv,GarciaEchevarria:2011rb,Echevarria:2012pw,%
Collins:2012uy} (see Fig.~\ref{f:trans_momenta_SIDIS} for a graphical
representation of the involved transverse momenta):  
\begin{align}
\label{e:SIDISkT}
   F_{UU,T}(x,z, \bm{P}_{h \Tperp}^2, Q^2) &= \sum_a \mathcal{H}_{UU,T}^{a}(Q^2) \\ 
      &\times x \int d^2\bm{k}_\T^{} \, d^2\bm{P}_\T^{} \,  f_1^a\big(x,\bm{k}_{\T}^2; Q^2 \big) \, D_{1}^{a\smarrow h}\big(z,\bm{P}_{\T}^2; Q^2 \big) \,
      \delta^{(2)} \big(z {\bm k}_{\T} - {\bm P}_{h \Tperp} + {\bm P}_{\T}\big)
\nonumber\\&
\nonumber + Y_{UU,T}\big(Q^2, \bm{P}_{h\Tperp}^2\big) + \mathcal{O}\big(M^2/Q^2\big) \, .
\end{align} 
Here, $\mathcal{H}_{UU,T}$ is the hard scattering part; $f_1^a(x,\bm{k}_{\T}^2;Q^2)$ is the TMD PDF of unpolarized partons with flavor $a$ in an unpolarized
proton, carrying longitudinal momentum fraction $x$ and transverse momentum
$\bm{k}_\T$.  The $D_1^{a\smarrow h}(z, \bm{P}_{\T}^2;Q^2)$ is the TMD FF describing the fragmentation of an unpolarized parton with flavor $a$ into
an unpolarized hadron $h$ carrying longitudinal momentum fraction $z$ and
transverse momentum 
$\bm{P}_\T$ (see Fig.~\ref{f:trans_momenta_SIDIS}).  
TMDs generally depend on two energy scales~\cite{Collins:2011zzd}, which enter
via the renormalization of ultraviolet and rapidity divergencies. 
In this work we choose them to be equal and set them to $Q^2$. 
The term $Y_{UU,T}$ is introduced to ensure a matching
to the perturbative fixed-order calculations at higher transverse momenta. 

In our analysis, we neglect any correction of the order of $M^2/Q^2$ or higher
to  Eq.~\eqref{e:SIDISkT}.
At large $Q^2$ this is well justified. 
However, fixed-target DIS experiments typically 
collect a large amount of data
at relatively low $Q^2$ values, where these assumptions
should be all tested in future studies. The reliability
of the theoretical description of SIDIS at low $Q^2$ has been recently
discussed in Refs.~\cite{Boglione:2016bph,Moffat:2017sha}.
 
Eq.~\eqref{e:SIDISkT} can be expanded in powers
of $\alpha_S$. In the present analysis, we 
will consider only the terms at order $\alpha_S^0$. In this case 
$\mathcal{H}^a_{UU,T} (Q^2) \approx e_a^2$
and $Y_{UU,T}\approx 0$. 
However, perturbative corrections include large logarithms $L \equiv
\log\big(z^2 Q^2/P_{hT}^2\big)$, so that $\alpha_S L \approx 1$.
In the present analysis, we will take into account all
leading and Next-to-Leading Logarithms (NLL).\footnote{We
  remark that formulas at NNLL are available in the
  literature~\cite{Echevarria:2016scs}.}  

In these approximations ($\alpha_S^0$ and NLL), only the first term in
Eq.~\eqref{e:SIDISkT} is relevant (often in the
literature this has been called $W$ term). We expect this term to provide a
good description of the
structure function only in the region where $P_{hT}^2 \ll Q^2$. 
It can happen that $Y_{UU,T}$, defined
in the standard way (see, e.g., Ref.~\cite{Collins:1984kg}), gives large
contributions also in this region, but it is admissible to
redefine it in order to avoid this problem~\cite{Collins:2016hqq}. 
We leave a detailed treatment of the matching to the high $P_{hT}^2 \approx Q^2$
region to future investigations.   

To the purpose of applying TMD evolution equations, we
 need to calculate the Fourier transform of the part of 
Eq.~\eqref{e:SIDISkT} involving TMDs. The structure function thus reduces to 
\begin{align}
\label{e:SIDISkTFF}
   F_{UU,T}(x,z, \bm{P}_{h \Tperp}^2, Q^2) &\approx 2\pi \sum_a e_a^2 x 
       \int_0^{\infty} {d \bT} \bT J_0\big(\bT |\bm{P}_{hT}|/z\big)
      \tilde{f}_1^a\big(x, \bT^2;Q^2\big) \tilde{D}_1^{a\smarrow h}\big(z, \bT^2;
      Q^2 \big) 
 \, .
\end{align} 
where we introduced the Fourier transforms of the TMD PDF and FF according to
\begin{align} 
\tilde{f}_1^a\big(x, \bT^2;Q^2\big) &=
\int_0^{\infty} d |\bm{k}_\T| 
                |\bm{k}_\T|J_0\big(\bT |\bm{k}_\T|\big) 
       f_1^a\big(x, \bm{k}_\T^2;Q^2\big),
\\
\tilde{D}_1^{a\smarrow h}\big(z, \bT^2; Q^2 \big) &=
\int_0^{\infty} \frac{d |\bm{P}_{\T}|}{z^2} |\bm{P}_{\T}| 
                                             J_0\big(\bT |\bm{P}_{\T}|/z\big)
       D_1^{a\smarrow h}\big(z, \bm{P}_{\T}^2; Q^2 \big).
\end{align}

\subsection{Drell--Yan and Z production}
\label{ss:DY_formalism}

\begin{figure}
\centering
\includegraphics[width=0.6\textwidth]{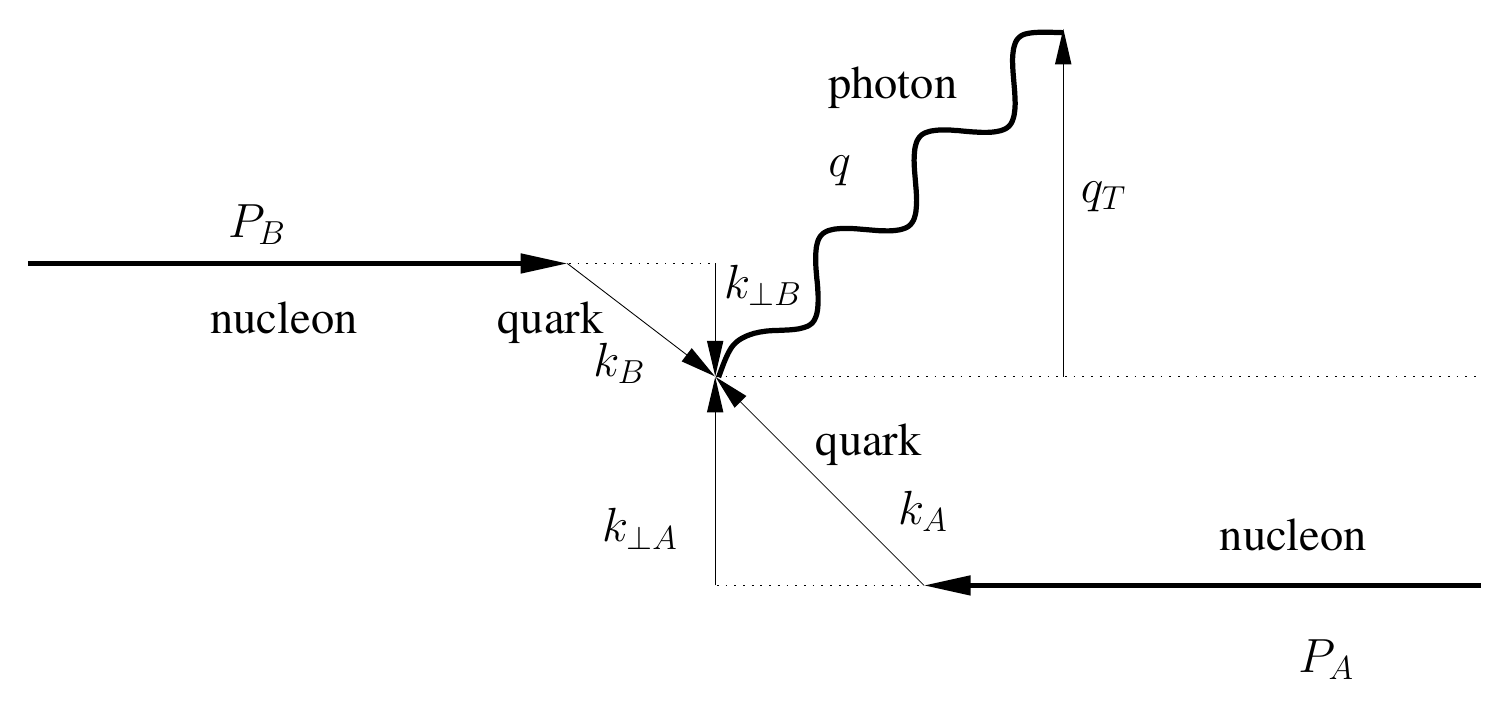}
\caption{Diagram describing the relevant momenta involved in a Drell--Yan
  event: two partons from two hadrons collide. They have transverse momenta
  $\bm{k}_{\T A}$ and  $\bm{k}_{\T B}$ (not measured). They produce a virtual
  photon with (measured) transverse momentum $\bm{q}_T=\bm{k}_{\T A}+\bm{k}_{\T B} $ with respect to the hadron
  collision axis.}
\label{f:trans_momenta_DY}
\end{figure}

In a Drell--Yan process, two hadrons $A$ and $B$ with momenta $P_A$ and $P_B$
collide at a center-of-mass energy squared $s = (P_A + P_B)^2$ 
and produce a virtual photon or a $Z$ boson plus hadrons. 
The boson decays into a
lepton-antilepton pair. The reaction formula is
\begin{equation}
A(P_A)+B(P_B)\to [\gamma^*/Z + X \to] \ell^+(l) + \ell^-(l') + X.
\end{equation} 
The invariant mass of the virtual photon is $Q^2=q^2$ with $q = l + l'$. 
We introduce the rapidity of the virtual photon/Z boson
\begin{equation}
\eta=\frac{1}{2}\log\bigg(\frac{q^0+q_z}{q^0-q_z}\bigg)\  .
\end{equation} 
where the $z$ direction is defined along the momentum of hadron A (see Fig.~\ref{f:trans_momenta_DY}).

The cross section can be written in terms of structure
functions~\cite{Boer:2006eq,Arnold:2008kf}. For our purposes, we need the unpolarized 
cross section
integrated over $d\Omega$ and over the azimuthal angle of the virtual photon, 
\begin{align}
\label{e:dsigma_gZ}
\frac{d\sigma}{dQ^2\, dq_T^2\,d\eta} &= \sigma_0^{\gamma,Z}
\bigg(F_{UU}^1 + \frac{1}{2} F_{UU}^2\bigg). 
\end{align} 
The elementary cross sections are
\begin{align}
\sigma_0^{\gamma} &= \frac{4\pi^2 \alpha^2_{\rm em}}{3 Q^2 s},
&
\sigma_0^Z &= 
\frac{\pi^2 \alpha_{\rm em}}{s \sin^2{\theta_W} \cos^2{\theta_W}}
B_R(Z\rightarrow \ell^+\ell^-)
\delta(Q^2 - M_Z^2), 
\label{e:elem_cs_sig0}
\end{align} 
where $\theta_W$ is Weinberg's angle, $M_Z$ is the mass of the $Z$ boson, and
$B_R(Z\rightarrow \ell^+\ell^-)$ 
is the branching ratio for the $Z$ boson decay in two leptons.
We adopted the narrow-width approximation, i.e., we neglect contributions for 
$Q^2 \neq M_Z^2$. 
We used the values 
$\sin^2 \theta_W= 0.2313$, $M_Z = 91.18$ GeV, and 
$B_R(Z\rightarrow \ell^+\ell^-)=3.366$~\cite{Olive:2016xmw}.  
Similarly to the SIDIS case, in the kinematic limit $q_T^2 \ll Q^2$ 
the structure function $F_{UU}^2$ can be neglected 
(for measurement and estimates of this
structure function see, e.g.,
Ref.~\cite{Lambertsen:2016wgj} and references therein). 

The longitudinal momentum fractions of the annihilating quarks
can be written in terms of
rapidity in the following way 
\begin{align}
x_A &= \frac{Q}{\sqrt{s}} e^{\eta},
&
x_B &= \frac{Q}{\sqrt{s}} e^{-\eta}.
\label{xab}
\end{align} 
Some experiments use the variable $x_F$, which is connected to the other
variables  by the following relations
\begin{align}
\label{e:eta_xf}
\eta &= \sinh^{-1}\bigg(\frac{\sqrt{s}}{Q}\frac{x_F}{2}\bigg),
& 
x_{A} &= \sqrt{\frac{Q^2}{s} + \frac{x_F^2}{4}} + \frac{x_F}{2},
&
x_B &= x_A - x_F.  
\end{align} 

The structure function $F_{UU}^1$ can be written as (see Fig.~\ref{f:trans_momenta_DY} for a graphical
representation of the involved transverse momenta)
\begin{align}
\label{e:DYkT}
   F_{UU}^1(x_A,x_B, \bm{q}_{T}^2, Q^2) &= \sum_a \mathcal{H}_{UU}^{1 a}(Q^2) \\ 
      &\times x_A x_B \int d^2\bm{k}_{\T A}^{} \, d^2\bm{k}_{\T B}^{} 
\,  f_1^a\big(x_A,\bm{k}_{\T A}^2; Q^2 \big) 
\, f_{1}^{\bar{a}}\big(x_B,\bm{k}_{\T B}^2; Q^2 \big) \,
      \delta^{(2)} \big({\bm k}_{\T A} - {\bm q}_T + {\bm k}_{\T B}\big)
\nonumber\\&
\nonumber + Y_{UU}^1\big(Q^2, \bm{q}_T^2\big) + \mathcal{O}\big(M^2/Q^2\big) \, .
\end{align} 

As in the SIDIS case, in our analysis we neglect the $Y_{UU}$ term
and we consider the hard coefficients only up to leading order in
the couplings, i.e.,
\begin{align} 
{\cal H}_{UU, \gamma}^{1 a}(Q^2) &\approx \frac{e_a^2}{N_c},
&
{\cal H}_{UU, Z}^{1 a}(Q^2) &\approx \frac{V_a^2+A_a^2}{N_c} \ ,
\end{align}  
where\footnote{We remind the reader that the value of weak isospin $I_3$ is equal to $+1/2$ for $u$, $c$, $t$ and
  $-1/2$ for $d$, $s$, $b$.}
\begin{align}
V_a & = I_{3a} - 2 e_{a} \sin^2 \theta_W \  ,
&
A_a & = I_{3a} \  .
\end{align} 

The structure function can be conveniently expressed as a Fourier transform of
the right-hand side of Eq.~\eqref{e:DYkT} as 
\begin{align}
\label{e:DYkTFF}
   F_{UU}^1(x_A,x_B, \bm{q}_T^2, Q^2) &\approx
 2\pi \sum_a {\cal H}_{UU}^{1 a} \,x_a x_B \int_0^{\infty} d \bT \bT\, J_0\big( \bT |\bm{q}_T|\big)\ 
      \tilde{f}_1^a\big(x_A, \bT^2;Q^2\big) \   \tilde{f}_1^{\bar{a}}\big(x_B, \bT^2;Q^2 \big)  \, .
\end{align}

\subsection{TMDs and their evolution}
\label{ss:TMDevo}

Evolution equations quantitatively describe the connection between different values for the energy scales. 
In the following we will set their initial values to $\mu_b^2$ and their final values as $Q^2$, so that only $Q^2$ and $\mu_b^2$ need to be specified in a TMD distribution.
Following the formalism of Refs.~\cite{Collins:2011zzd,Aybat:2011zv}, the
unpolarized TMD distribution and fragmentation functions in configuration
space for a parton with flavor $a$ at a certain scale $Q^2$ can be written as 
\begin{align}   
\widetilde{f}_1^a (x,  \bT^2; Q^2) &= \sum_{i=q,\bar q,g} \bigl( C_{a/i} \otimes f_1^i \bigr) (x,\bar{\bb}_\ast,\mu_b^2) 
\  e^{S (\mu_b^2, Q^2)} \bigg(
\frac{Q^2}{\mu_b^2}\bigg)^{-K(\bar{\bb}_\ast;\mu_b)}
\  
\bigg(\frac{Q^2}{Q_0^2} \bigg)^{g_K(\bT)} \  \widetilde{f}_{1 {\rm NP}}^a (x, \bT^2) \ ,
\label{e:TMDevol1} \\
\widetilde{D}_1^{a\to h} (z, \bT^2; Q^2) &= \sum_{i=q,\bar q,g} \bigl(
\hat{C}_{a/i} \otimes D_1^{i\to h} \bigr) (z,\bar{\bb}_\ast, \mu_b^2) \  e^{S
  (\mu_b^2, Q^2)} \bigg(\frac{Q^2}{\mu_b^2}\bigg)^{-K(\bar{\bb}_\ast;\mu_b)}
\  
\bigg(\frac{Q^2}{Q_0^2} \bigg)^{g_K( \bT)} \  \widetilde{D}_{1 {\rm NP}}^{a\to h} (z, \bT^2) \  .
\label{e:TMDevol2}
\end{align}

We choose the scale $\mu_b$ to be
\begin{align} 
\mu_b &= \frac{2 e^{-\gamma_E}}{\bar{\bb}_{\ast}} \  ,
\label{e:mub}
\end{align}  
where $\gamma_E$ is the Euler constant and
\begin{align} 
\bar{\bb}_{\ast} &\equiv \bar{\bb}_{\ast}(\bT;\bb_{\rm min},\bb_{\rm max}) = \bb_{\rm max} \Bigg(\frac{1-e^{- \bT^4 / \bb_{\rm max}^4} }{1-e^{- \bT^4 / \bb_{\rm min}^4}} \Bigg)^{1/4} .
\label{e:b*}
\end{align}  
This variable replaces the simple dependence upon $\bT$ 
in the perturbative parts of the
  TMD definitions of Eqs.~\eqref{e:TMDevol1},~\eqref{e:TMDevol2}. In fact, at
  large $\bT$ these parts are no longer reliable. Therefore, the
  $\bar{\bb}_{\ast}$ is chosen to saturate on the maximum value $\bb_{\rm max}$,
  as suggested by the CSS 
  formalism~\cite{Collins:2011zzd,Aybat:2011zv}.
On the other hand, at
small $\bT$ the TMD formalism is not valid and should be 
matched to the fixed-order collinear
calculations. The way
the matching is implemented is not unique.  In any case, the TMD contribution
can be arbitrarily modified at small $\bT$. In our approach, we choose to
saturate 
$\bar{\bb}_{\ast}$  at
the minimum value $\bb_{\rm min}\propto 1/Q$. With the appropriate choices, 
for $\bT=0$ the Sudakov exponent vanishes, as it
should~\cite{Parisi:1979se,Altarelli:1984pt}. 
Our choice partially corresponds to modifying the resummed logarithms as in
Ref.~\cite{Bozzi:2010xn} and to other similar modifications proposed in the
literature~\cite{Boer:2014tka,Collins:2016hqq}. One advantage of these kind of
prescriptions is that by integrating over the impact parameter $\bT$, the
collinear expression for the cross section in terms of collinear PDFs is
recovered, at least at leading order~\cite{Collins:2016hqq}.
We remind the reader that there are different schemes available to deal with
the high-$\bT$ region, such as the the so-called ``complex-$\bb$
prescription''~\cite{Laenen:2000de} or an extrapolation of the perturbative
small-$\bb_T$ calculation to the large $\bb_T$ region based on dynamical power
corrections~\cite{Qiu:2000hf}. 

The values of $\bb_{\rm max}$ and
$\bb_{\rm min}$ could
be regarded as arbitrary scales separating perturbative from nonperturbative
regimes. 
We choose to fix them to the
values 
\begin{align}
\bb_{\rm max} &= 2 e^{-\gamma_E}  \text{  GeV}^{-1} \approx 1.123 \text{  GeV}^{-1}, 
&
\bb_{\rm min} &= 2 e^{-\gamma_E}/Q \ .
\label{e:bminmax}
\end{align} 
The motivations are the following: 
\begin{itemize}
\item{} with the above choices, the scale $\mu_b$ is
  constrained between 1 GeV and $Q$, so that the collinear PDFs are never
  computed at a scale lower than 1 GeV and the lower limit of the integrals
  contained in the definition of the perturbative Sudakov factor (see
  Eq.~\eqref{e:Sudakov})  can never
  become larger than the upper limit;
\item{} at $Q=Q_0 = 1$ GeV, $\bb_{\rm max} = \bb_{\rm min}$ and there are no evolution effects; the TMD is
simply given by the corresponding collinear function multiplied by a
nonperturbative contribution depending on $k_\T$ (plus possible corrections of
order $\alpha_S$ from the Wilson coefficients).
\end{itemize} 

At NLL accuracy, for our choice of scales $K(\bar{\bb}_{\ast},\mu_b) =
0$. 
Similarly, the $C$ and $\hat{C}$ are perturbatively calculable Wilson
coefficients for 
the TMD distribution and fragmentation functions, respectively. They are
convoluted with the corresponding collinear functions according to 
\begin{align}
\bigl( C_{a/i} \otimes f_1^i \bigr) (x, \bar{\bb}_\ast, \mu_b^2) &=
  \int_x^1 \frac{du}{u}\  
        C_{a/i} \Big( \frac{x}{u}, \bar{\bb}_\ast, \alpha_S\big(\mu_b^2\big)  \Big) \  
        f_1^i (u; \mu_b^2) \  , 
\label{e:WC1} \\
\bigl( \hat{C}_{a/i} \otimes D_1^{i\to h} \bigr) (z, \bar{\bb}_\ast, \mu_b^2) &= \int_z^1 \frac{du}{u}\  \hat{C}_{a/i} \left( \frac{z}{u}, \bar{\bb}_\ast, \alpha_S\big(\mu_b^2\big) \right) \  D_1^{i\to h} (u; \mu_b^2) \  . 
\label{e:WC2}
\end{align}

In the present analysis, we consider only the leading-order term
in the $\alpha_S$ expansion for $C$ and $\hat{C}$ , i.e., 
\begin{align} 
C_{a/i} \Big( \frac{x}{u}, \bar{\bb}_\ast, \alpha_S\big(\mu_b^2\big)  \Big) 
&\approx
\delta_{ai} \delta(1-x/u),
&
\hat{C}_{a/i} \Big( \frac{z}{u}, \bar{\bb}_\ast, \alpha_S\big(\mu_b^2\big) \Big)
 &\approx
\delta_{ai} \delta(1-z/u).
\end{align}  
As a consequence of the choices we made, 
the expression for the evolved TMD functions reduces to
\begin{align}   
\widetilde{f}_1^a (x,  \bT^2; Q^2) &= f_1^a (x ; \mu_b^2) 
\  e^{S (\mu_b^2, Q^2)} \  e^{g_K(\bT) \ln (Q^2 / Q_0^2)} \  \widetilde{f}_{1 {\rm NP}}^a (x, \bT^2) \ ,
\label{e:TMDevol1b} \\
\widetilde{D}_1^{a\to h} (z, \bT^2; Q^2) &= D_1^{a\to h} (z; \mu_b^2) \  e^{S (\mu_b^2, Q^2)} \  e^{g_K( \bT) \ln (Q^2 / Q_0^2)} \  \widetilde{D}_{1 {\rm NP}}^{a\to h} (z, \bT^2) \  .
\label{e:TMDevol2b}
\end{align}

The Sudakov exponent $S$ 
can be written as
\begin{equation} 
S(\mu_b^2,Q^2)=-\int_{\mu_b^2}^{Q^2}{d\mu^2\over \mu^2}
\bigg[A\Big(\alpha_S(\mu^2)\Big)\ln\bigg({Q^2\over \mu^2}\bigg) 
+ B\Big(\alpha_S(\mu^2)\Big) \bigg] \ ,
\label{e:Sudakov} 
\end{equation} 
where the functions $A$ and $B$ have a perturbative expansions of the form
\begin{align}
A\left(\alpha_S(\mu^2)\right) &= \sum_{k=1}^{\infty}A_k
\bigg(\frac{\alpha_S}{\pi} \bigg)^k,
&
B\left(\alpha_S(\mu^2)\right) &= \sum_{k=1}^{\infty}B_k
\bigg(\frac{\alpha_S}{\pi} \bigg)^k.
\end{align} 
To NLL accuracy, we need the following 
terms~\cite{Davies:1984hs,Collins:1984kg}
\begin{align}
A_1&= C_F, 
&
A_2&=
\frac{1}{2} C_F  \bigg[
C_A \bigg( \frac{67}{18} - \frac{\pi^2}{6} \bigg)
- \frac{5}{9} N_f \bigg],
&
B_1&= - \frac{3}{2}C_F.
\end{align} 
We use the approximate analytic expression for $\alpha_S$ at NLO with the
$\Lambda_{\text{QCD}}=$ 340 MeV, 296 MeV, 214 MeV for three, four, five
flavors, respectively, corresponding to a value of $\alpha_S(M_Z)=0.117$. 
We fix the flavor thresholds at $m_c=1.5$ GeV and 
$m_b= 4.7$ GeV. The integration of the Sudakov exponent in
Eq.~\eqref{e:Sudakov} can
be done analytically (for the complete expressions see, e.g.,
Refs.~\cite{Frixione:1998dw,Bozzi:2005wk,Echevarria:2012pw}).  

Following Refs.~\cite{Nadolsky:1999kb,Landry:2002ix,Konychev:2005iy}, for the
nonperturbative Sudakov factor we make the traditional choice 
\begin{equation}
g_K (\bT) = - g_2 \bT^2 / 2
\end{equation} 
with $g_2$ a free parameter. Recently, several alternative
forms have been proposed~\cite{Aidala:2014hva,Collins:2014jpa}. Also, recent
theoretical studies aimed at calculating this term using nonperturbative
methods~\cite{Scimemi:2016ffw}.  
All these
choices should be tested in future studies. In Ref.~\cite{D'Alesio:2014vja}, a
good agreement with data was achieved even without this term, but this is
not possible when including data at low $Q^2$.

In this analysis, for the collinear PDFs $f_1^a$ we adopt the GJR08FFnloE
set~\cite{Gluck:2007ck} through the LHAPDF library~\cite{Buckley:2014ana}, and
for the collinear fragmentation functions 
the DSS14 NLO set for
pions~\cite{deFlorian:2014xna} and the DSS07 NLO set for
kaons~\cite{deFlorian:2007aj}.\footnote{After the completion of our analysis, 
a new set of kaon
  fragmentation function was presented in Ref.~\cite{deFlorian:2017lwf}.} We
will comment on the use of other PDF sets in Sec.~\ref{ss:replica105}.

We parametrize the intrinsic nonperturbative parts of the TMDs in the
following ways
\begin{align}
\widetilde{f}_{1 {\rm NP}}^a (x, \bT^2) &= \frac{1}{2\pi}
        e^{-g_{1a} \frac{\bT^2}{4}}
        \bigg( 1 - \frac{\lambda  g_{1a}^{2}}{1+\lambda g_{1a}}  \frac{\bT^2}{4} \bigg)\  ,
\label{e:f1NP} \\
\widetilde{D}_{1 {\rm NP}}^{a \to h} (z, \bT^2) &= 
    \frac{ g_{3 a\to h} \   e^{-g_{3 a\to h} \frac{\bT^2}{4 z^2}}
        + \big(\lambda_F/z^2\big)    g_{4 a\to h}^{2}
    \left(1 - g_{4 a\to h} \frac{\bT^2}{4 z^2} \right)
         \  e^{- g_{4 a\to h}^{2}  \frac{\bT^2}{4z^2}}}
     {2 \pi z^2 \Big(g_{3 a\to h} + \big(\lambda_F/z^2\big)    g_{4 a\to h}^{2}\Big)} \  .
\label{e:D1NP}
\end{align} 
After performing the anti-Fourier transform, the $f_{1 \rm NP}$ and $D_{1\rm
  NP}$ in momentum space correspond to 
\begin{align} 
f_{1 {\rm NP}}^a (x, \bm{k}_{\T}^2) &= \frac{1}{\pi} \  
                        \frac{\big( 1 +\lambda \bm{k}_{\T}^2\big)}
                                { g_{1a} +\lambda \   g_{1a}^2}
                        \  e^{- \frac{\bm{k}_{\T}^2}{g_{1a}}} \  ,
\label{e:f1NPk}   \\
D_{1 {\rm NP}}^{a\to h} (z, \bm{P}_{\T}^2) &=  \frac{1}{\pi} \   
                  \frac{1}{g_{3 a\to h} +
                    \big(\lambda_F/z^2\big) g_{4 a \to h}^{2}}
           \   \bigg( e^{- \frac{\bm{P}_{\T}^2}{g_{3 a \to h}}}
                            + \lambda_F \frac{\bm{P}_{\T}^2}{z^2} \  
           e^{- \frac{\bm{P}_{\T}^2}{g_{4 a \to h}}} \bigg) \  .
\label{e:D1NPk}
\end{align} 
The TMD PDF at the starting scale is therefore a normalized sum of a Gaussian
with variance $g_1$ and the same Gaussian weighted by a factor $\lambda
\bm{k}_{\T}^2$. The TMD FF at the starting scale is a normalized sum of a Gaussian
with variance $g_3$ and a second Gaussian with variance $g_4$ weighted by a factor $\lambda_F
\bm{P}_{\T}^2/z^2$. 
The choice of this particular functional forms is motivated by model
calculations: the weighted Gaussian in the TMD PDF could arise from 
the presence of components of the quark wave function 
with angular momentum
$L=1$~\cite{Bacchetta:2007wc,Pasquini:2008ax,Avakian:2010br,Bacchetta:2010si,Burkardt:2015qoa}. Similar
features occur in models of fragmentation functions~\cite{Bacchetta:2002tk,Bacchetta:2007wc,Matevosyan:2011vj}.

The
Gaussian width of the TMD distributions may depend 
on the parton flavor $a$~\cite{Signori:2013mda,Matevosyan:2011vj,Schweitzer:2012hh}. 
In the present analysis, however, we assume they are
flavor independent. The justification for this choice is that most of the data
we are considering are not sufficiently sensitive to flavor differences,
leading to unclear results. We
will devote attention to this issue in further studies.

Finally, we assume that the Gaussian width of the TMD depends on the 
fractional longitudinal momentum $x$ according to
\begin{align} 
 g_1 (x) &= N_1 \;  
\frac{(1-x)^{\alpha} \  x^{\sigma} }{ (1 - \hat{x})^{\alpha} \  \hat{x}^{\sigma} } \, ,
\label{e:kT2_kin}
\end{align}
where $\alpha, \, \sigma,$ and $N_1 \equiv  g_1 (\hat{x})$ with $\hat{x} = 0.1$,
are free parameters. Similarly, for fragmentation functions we have
\begin{align}  
g_{3,4} (z) &= N_{3,4} \  
               \frac{ (z^{\beta} + \delta)\ (1-z)^{\gamma} }{ (\hat{z}^{\beta} + \delta)\   (1 - \hat{z})^{\gamma} } \, ,
 \label{e:PT2_kin}
 \end{align}
where $\beta, \, \gamma, \, \delta, $ and $N_{3,4} \equiv g_{3,4} (\hat{z})$ with
$\hat{z} = 0.5$ 
are free parameters. 

The average transverse momentum squared for the distributions in
Eq.~\eqref{e:f1NPk} and \eqref{e:D1NPk} can be computed analytically:
\begin{align}
\big \langle \bm{k}_{\perp}^2 \big \rangle (x) &= \frac{g_1(x) + 2 \lambda g_1^2(x)}
{1+ \lambda g_1(x)},
&
\big \langle \bm{P}_{\perp}^2 \big \rangle (z) &= \frac{g_3^2(z) + 2 \lambda_F
  g_4^3(z)}{g_3(z) + \lambda_F g_4^2(z)}.
\label{e:transmom2}
\end{align} 

\section{Data analysis}
\label{s:data_analysis}

The main goals of our work are to extract information about intrinsic transverse momenta, to study the evolution of TMD parton distributions and fragmentation functions over a large enough range of energy, and to test their universality among different processes.
To achieve this we included measurements taken from SIDIS, Drell--Yan and $Z$
boson production from different experimental collaborations at different
energy scales. 
In this chapter we describe the data sets considered for each process and the applied kinematic cuts.

Tab.~\ref{t:data_SIDIS_proton} refers to the data sets for SIDIS off proton target (\hermes\ experiment) and presents their kinematic ranges. 
The same holds for Tab.~\ref{t:data_SIDIS_deuteron}, Tab.~\ref{t:data_DY}, Tab.~\ref{t:data_Z} for SIDIS off deuteron (\hermes\ and \compass\ experiments), Drell--Yan events at low energy and $Z$ boson production respectively. 
If not specified otherwise, the theoretical formulas are computed at the
average values of the kinematic variables in each bin.

\subsection{Semi-inclusive DIS data}
\label{ss:SIDIS data}

The SIDIS data are taken from \hermes~\cite{Airapetian:2012ki} and \compass~\cite{Adolph:2013stb} experiments. 
Both data sets have already been analyzed in previous works,
e.g., Refs.~\cite{Signori:2013mda,Anselmino:2013lza}, 
however they have never been
fitted together, including also the contributions deriving from TMD
evolution. 

The application of the TMD formalism to SIDIS depends on the capability of
identifying the current fragmentation region. This task has been recently
discussed in Ref.~\cite{Boglione:2016bph}, where the authors point out a
possible overlap among different  fragmentation regions when the hard scale
$Q$ is sufficiently low.  
In this paper we do not tackle this problem and we leave it to future
studies. As described in Tabs.~\ref{t:data_SIDIS_proton}
and~\ref{t:data_SIDIS_deuteron}, we identify the current fragmentation region
operating a cut on $z$ only, namely $0.2 < z < 0.74$.

Another requirement for the applicability of TMD factorization is the presence
of two separate scales in the process. In SIDIS, those are the $Q^2$ and
$P_{hT}^2$, which should satisfy the condition $P_{hT}^2 \ll Q^2$, or more
precisely  $P_{hT}^2/z^2 \ll Q^2$. 
We implement this condition by imposing 
$P_{hT} < \min[0.2\ Q, 0.7\ Qz] + 0.5$ GeV.
With this choice, $P_{hT}^2$ is always smaller than $Q^2/3$, but in a few 
bins (at low $Q^2$ and $z$) $P_{hT}^2/z^2$ may become larger
than $Q^2$. The applicability of TMD factorization in this case could be
questioned. 
However, as we will explain further in Sec.~\ref{ss:replica105},
we can obtain a fit that can describe a wide region of $P_{hT}$ and can also
perform very well in a restricted region,  where TMD factorization
certainly holds.

All these choices are summarized in Tabs.~\ref{t:data_SIDIS_proton} and~\ref{t:data_SIDIS_deuteron}.

\subsubsection{\hermes\ data}
\label{sss:hermes}

\hermes\ hadron multiplicities are measured in a fixed target experiment,
colliding a $27.6$ GeV lepton beam on a hydrogen ($p$) or deuterium ($D$) gas
target, for a total of 2688 points.
These are grouped in bins of $(x,z,Q^2,P_{hT})$ with the average values of $(x,Q^2)$ ranging from about $(0.04, 1.25\text{ GeV}^2)$ to $(0.4, 9.2\text{ GeV}^2)$. 
The collinear energy fraction $z$ in Eq.~\eqref{e:xyz} ranges in $0.1\leq z\leq 0.9$. The transverse momentum of the detected hadron satisfies $0.1 \text{ GeV} \leq \vert P_{hT} \vert \leq 1.3 \text{ GeV}$.
The peculiarity of \hermes\ SIDIS experiment lies in the ability of its detector to distinguish between pions and kaons in the final state, in addition to determining their momenta and charges.
We consider eight different 
combinations of target ($p,\,D$) and detected
charged hadron ($\pi^\pm, \,  K^\pm$ ). The
\hermes\ collaboration published two distinct sets, characterized by the inclusion or subtraction of the vector meson contribution. In our work we considered only the data set where this contribution has been subtracted.

\subsubsection{\compass\ data}
\label{sss:compass}

The \compass\ collaboration extracted multiplicities for charge-separated but
unidentified hadrons produced in SIDIS off a deuteron ($^6\text{LiD}$)
target~\cite{Adolph:2013stb}.  The number of data points is an order of magnitude higher compared to the \hermes\ experiment.
The data are organized in multidimensional bins of $(x,z,Q^2,P_{h\Tperp})$,
they cover a range in $(x,Q^2)$ from about $(0.005, 1.11\text{ GeV}^2)$ to
$(0.09, 7.57\text{ GeV}^2)$ and the interval $0.2 \leq z \leq 0.8$. 
The multiplicities published by \compass\ are affected by normalization
errors (see the {\em erratum} to Ref.~\cite{Adolph:2013stb}). In order to
avoid this issue, we divide the data in each bin in ($x, z, Q^2$) by the data
point with the lowest $P_{hT}^2$ in the bin. 
As a result, we define the {\em normalized} multiplicity as
\begin{equation}
m_{\text{norm}}(x,z,\bm{P}_{h\Tperp}^2, Q^2) = \frac{m_N^h(x,z,\bm{P}_{h\Tperp}^2, Q^2)}{m_N^h (x,z,{\rm \min}[\bm{P}_{h\Tperp}^2], Q^2)} \ ,
\label{e:mult_norm}
\end{equation}
where the multiplicity $m_N^h$ is defined in Eq.~\eqref{e:multiplicity}. When
fitting normalized multiplicities, the first data point of each bin is
considered as a fixed constraint and excluded from the degrees of freedom.

\subsection{Low-energy Drell--Yan data}
\label{ss:dy}

We analyze Drell--Yan events collected by fixed-target experiments at
low-energy. These data sets have been considered also in previous works, e.g.,
in Ref.~\cite{Landry:1999an,Landry:2002ix,Konychev:2005iy,DAlesio:2014mrz}. 
We used data sets from the E288 experiment~\cite{Ito:1980ev}, which measured the invariant dimuon cross section $E d^3\sigma / dq^3$ for the production of $\mu^+ \mu^-$ pairs from the collision of a proton beam with a fixed target, either composed of Cu or Pt.
The measurements were performed using proton incident energies of $200$, $300$ and $400$ GeV, producing three different data sets.
Their respective center of mass energies are $\sqrt{s}=19.4,23.8,27.4$ GeV.
We also included the set of measurements $E d^3\sigma / dq^3$ from E605~\cite{Moreno:1990sf}, extracted from the collision of a proton beam with an energy of $800$ GeV ($\sqrt{s}=38.8$ GeV) on a copper fixed target .

The explored $Q$ values are higher compared to the SIDIS case, as can be seen
in Tab.~\ref{t:data_DY}. E288 provides data at fixed rapidity,
whereas E605 provides data at fixed $x_F=0.1$.  
We can apply TMD factorization if
$q_T^2 \ll Q^2$, where $q_T$ is the transverse
momentum of the intermediate electroweak boson, reconstructed from the
kinematics of the final state leptons. We choose $q_T <
0.2\ Q + 0.5$ GeV. 
As suggested in Ref.~\cite{Ito:1980ev}, we consider the target nuclei as an
incoherent ensemble composed $40\%$ by protons and $60\%$ by neutrons.

As we already observed, results from E288 and E605 experiments are reported as
$\frac{Ed^3\sigma}{d^3q}$; this variable is related to the differential cross
section of Eq.~\eqref{e:dsigma_gZ} in the following way:
\begin{equation}
\frac{Ed^3\sigma}{d^3q}=\frac{d^3\sigma}{d\phi d\eta q_T dq_T} \Rightarrow \frac{d^2\sigma}{\pi d\eta d(q^2_T)},
\end{equation}
where $\phi$ is the polar angle of $q_T$ and 
the third term is the average over $\phi$.
Therefore, the invariant dimuon cross section can be obtained from Eq.~\eqref{e:dsigma_gZ} integrating over $Q^2$ and adding a factor $1/\pi$ to the result
\begin{equation}
\frac{Ed^3\sigma}{d^3q} = \frac{1}{\pi} \int dQ^2 \frac{d\sigma}{dQ^2dq^2_T d\eta} .
\end{equation}

Numerically we checked that integrating in $Q^2$ only the prefactor
$\sigma_q^\gamma$ (see Eq.~\eqref{e:elem_cs_sig0}) introduces only
a negligible error in the theoretical estimates. We also assume that
$\alpha_{\rm em}$ does not change within the experimental bin. Therefore, for Drell--Yan we
obtain
\begin{equation}
\frac{1}{\pi} \int dQ^2 \frac{d\sigma}{dQ^2dq^2_T d\eta}
\approx 
\frac{4\alpha_{\rm em}^2}{3 s} \ln \left( \frac{Q_f^2}{Q_i^2} \right)
F_{UU}^1. 
\end{equation}
where $Q_{i,f}$ are the lower and upper values in the experimental bin.

\subsection{Z-boson production data}
\label{ss:zboson}

In order to reach higher $Q$ and $q_T$ values, we also consider $Z$ boson production in collider experiments at Tevatron. 
We analyze data from CDF and D0, collected during Tevatron Run I~\cite{Affolder:1999jh,Abbott:1999wk} at $\sqrt{s}=1.8\text{ TeV}$ and Run II~\cite{Aaltonen:2012fi,Abazov:2007ac} at $\sqrt{s}=1.96\text{ TeV}$. CDF and D0 collaborations studied the differential cross section for the production of an $e^+e^-$ pair from $p\bar{p}$ collision through an intermediate $Z$ vector boson, namely $p\bar{p}\rightarrow Z \rightarrow e^+e^- + X$. 

The invariant mass distribution peaks at the $Z$-pole, $Q \approx M_Z$, while the transverse momentum of the exchanged $Z$ ranges in $0< q_T < 20 \text{ GeV}$.
We use the same kinematic cut applied to Drell--Yan events:  $q_T < 0.2\ Q + 0.5$ GeV $ = 18.7$ GeV, since $Q$ is fixed to $M_Z$. 

The observable measured in CDF and D0 is 
\begin{equation} 
\frac{d\sigma}{d q_T} = \int d Q^2 d \eta 2 q_T
\frac{d\sigma}{dQ^2dq^2_T d\eta}
\approx 
\frac{\pi^2 \alpha_{\rm em}}{s \sin^2{\theta_W} \cos^2{\theta_W}}
B_R(Z\rightarrow \ell^+\ell^-) 2 q_T \int d \eta F_{UU}^1,
\end{equation}  
apart from the case
of D0 Run II, for which the published data refer to $1/\sigma \times
d\sigma/dq_T$. In order to work with the same observable, we multiply the
D0 Run II data by the total cross section of the process $\sigma_{exp} = 255.8
\pm 16 \text{ pb}$~\cite{Abulencia:2005ix}. In this case, we add in quadrature
the uncertainties of the total cross section and of the published data. 

We normalize our functional form with the factors listed in
Tab.~\ref{t:data_Z}. These are the same normalization factors used in
Ref.~\cite{DAlesio:2014mrz}, computed by comparing the experimental total
cross section with the theoretical results based on the code of 
Ref.~\cite{Catani:2009sm}. These factors are not precisely consistent with our
formulas. In fact, as we will discuss in Sec.~\ref{ss:replica105} a 5\%
increase in these factors would improve the agreement with data, without
affecting the TMD parameters.

\renewcommand{\tabcolsep}{0.4pc} 
\renewcommand{\arraystretch}{1.3} 

\begin{table}[h!]
\begin{center}
\begin{tabular}{|c|c|c|c|c|}
 \hline
  & \hermes & \hermes & \hermes & \hermes \\
 ~          &  $p \to \pi^+$    &   $p \to \pi^-$    &  $p \to K^+$    &   $p \to K^-$               \\
 \hline
 Reference & \multicolumn{4}{c|}{\cite{Airapetian:2012ki}}        \\
\hline
\multirow{3}{*}{Cuts}             & \multicolumn{4}{c|}{$Q^2 > 1.4 \text{ GeV}^2$}     \\
             & \multicolumn{4}{c|}{$0.20 <z <0.74$}     \\
             & \multicolumn{4}{c|}{$P_{h \Tperp}< {\rm Min}[0.2\ Q, 0.7 \ Q z ] +0.5$ GeV}     \\
\hline
 Points         &  190 & 190 & 189 & 187       \\
 \hline
Max. $Q^2$      &  \multicolumn{4}{c|}{$9.2 \text{ GeV}^2 $}               \\
 \hline
$x$ range       & \multicolumn{4}{c|}{$0.04 < x < 0.4$ }                \\
\hline
\end{tabular}
\caption{SIDIS proton-target data (\hermes\ experiment).}
\label{t:data_SIDIS_proton}
\end{center}
\end{table}
\begin{table}[h!]
\begin{center}
\begin{tabular}{|c|c|c|c|c|c|c|}
 \hline
  & \hermes & \hermes & \hermes & \hermes & \compass & \compass\\
 ~          &  $D \to \pi^+$    &   $D \to \pi^-$    &  $D \to K^+$    &   $D \to K^-$      &  $D \to h^+$    &   $D \to h^-$            \\
 \hline
 Reference & \multicolumn{4}{c|}{\cite{Airapetian:2012ki}}        &\multicolumn{2}{c|}{\cite{Adolph:2013stb}} \\
\hline
\multirow{3}{*}{Cuts}             & \multicolumn{6}{c|}{$Q^2 > 1.4 \text{ GeV}^2$}     \\
             & \multicolumn{6}{c|}{$0.20 <z <0.74$}     \\
             & \multicolumn{6}{c|}{$P_{h \Tperp}< {\rm Min}[0.2\ Q, 0.7 \ Q z ] +0.5$ GeV}     \\
\hline
 Points         &  190 & 190 & 189 & 189   & 3125 & 3127   \\
 \hline
Max. $Q^2$      &  \multicolumn{4}{c|}{$9.2 \text{ GeV}^2 $}      & \multicolumn{2}{c|}{$10 \text{ GeV}^2 $}             \\
 \hline
$x$ range       & \multicolumn{4}{c|}{$0.04 < x < 0.4$ }    &  \multicolumn{2}{c|}{$0.005 < x < 0.12$ }             \\
\hline
Notes         &\multicolumn{4}{c|}{ }   & \multicolumn{2}{c|}{Observable: $\displaystyle m_{\text{norm}}(x,z,\bm{P}_{h\Tperp}^2, Q^2)$, Eq.~\eqref{e:mult_norm}}  \\
\hline 
\end{tabular}
\caption{SIDIS deuteron-target data (\hermes\ and \compass\ experiments).}
\label{t:data_SIDIS_deuteron}
\end{center}
\end{table}
\begin{table}[h!]
\begin{center}
\renewcommand{\tabcolsep}{0.4pc} 
\renewcommand{\arraystretch}{1.2} 
\begin{tabular}{|c|c|c|c|c|}
 \hline
 ~                        &  E288 200    &  E288 300        &  E288 400          &  E605                \\
 \hline
Reference               &  \cite{Ito:1980ev}  &   \cite{Ito:1980ev}  &  \cite{Ito:1980ev}  &   \cite{Moreno:1990sf}  \\
\hline
Cuts             & \multicolumn{4}{c|}{$q_T < 0.2\ Q +0.5$ GeV}
\\
 \hline
 Points                   &      45      &   45             &       78           &     35               \\
 \hline
 $\sqrt{s}$               &    19.4 GeV   &   23.8 GeV        &      27.4 GeV    &  38.8 GeV           \\
\hline
$Q$ range                 &  4-9 GeV      &  4-9 GeV         &  5-9, 11-14 GeV   &  7-9, 10.5-11.5 GeV   \\
 \hline
 Kin. var.           & $\eta$=0.40         &  $\eta$=0.21          &   $\eta$=0.03         &    $x_F=0.1$         \\
\hline
\end{tabular}
\caption{Low energy Drell--Yan data collected by the E288 and E605 experiments at Tevatron, with different center-of-mass energies.}
\label{t:data_DY}
\end{center}
\end{table}
\begin{table}[h!]
\begin{center}
\renewcommand{\tabcolsep}{0.4pc} 
\renewcommand{\arraystretch}{1.2} 
\begin{tabular}{|c|c|c|c|c|}
 \hline
 ~                        & CDF Run I    &  D0 Run I        & CDF Run II        & D0 Run II      \\
 \hline
 Reference        &\cite{Affolder:1999jh} &\cite{Abbott:1999wk}&\cite{Aaltonen:2012fi}&\cite{Abazov:2007ac} \\
\hline
Cuts             & \multicolumn{4}{c|}{$q_T< 0.2\ Q +0.5 \text{ GeV}=18.7$ GeV}                                  \\
\hline
 Points                   &      31      &   14             &       37          &        8       \\
 \hline
 $\sqrt{s}$               &      1.8 TeV &   1.8 TeV        &       1.96 TeV    &       1.96 TeV   \\
 \hline
Normalization        &  1.114       &    0.992          &       1.049        &       1.048    \\
\hline
\end{tabular}
\caption{$Z$ boson production data collected by the CDF and D0 experiments at Tevatron, with different center-of-mass energies.}
\label{t:data_Z}
\end{center}
\end{table}

\subsection{The replica method}
\label{ss:replica_method}

Our fit is based on the replica method. 
In this section we describe it and we give a definition of the $\chi^2$ function minimized by the fit procedure.
The fit and the error analysis are 
carried out using a similar Monte Carlo approach as in Refs.~\cite{Bacchetta:2012ty,Signori:2013mda,Radici:2015mwa} and taking
inspiration from the work of the Neural-Network PDF (NNPDF) collaboration
(see, e.g., Refs.~\cite{Forte:2002fg,Ball:2008by,Ball:2010de}). 
The approach consists in creating $\mathcal{M}$ replicas of the data points. In each replica (denoted by the index $r$), each data point $i$ is shifted by a Gaussian noise with the same variance as the measurement. 
Each replica, therefore, represents a possible outcome of an independent experimental measurement, which we denote by $m_{N, r}^{h}(x, z, \bm{P}_{h\Tperp}^2, Q^2)$. 
The number of replicas is chosen so that the mean and standard deviation of
the set of replicas accurately reproduces the original data points. In this
case 200 replicas are sufficient for the purpose. 
The error for each replica is taken to be equal to the error on the original data points. This is consistent with the fact that the variance of the $\mathcal{M}$ replicas should reproduce the variance of the original data points. 

A minimization procedure is applied to each replica separately, by minimizing the following error function: 
\begin{equation}
E_r^2(\{p\})=\sum_{i} 
\frac{\Bigl(m_{N, r}^{h}(x_i, z_i, \bm{P}_{h\Tperp i}^2, Q_i^2) - m_{N,  \mbox{\tiny theo}}^{h}(x_i, z_i, \bm{P}_{h\Tperp i}^2; \{p\})\Bigr)^2}
        {\Bigl( \Delta m_{N, \mbox{\tiny stat}}^{h\ 2} + \Delta m_{N, \mbox{\tiny sys}}^{h\ 2} \Bigr)(x_i, z_i, \bm{P}_{h\Tperp i}^2, Q^2_i) +\Bigl(\Delta m_{N, \mbox{\tiny theo}}^{h}(x_i, z_i, \bm{P}_{h\Tperp i}^2) \Bigr)^2}\  . 
\label{e:MC_chi2}
\end{equation}
The sum runs over the $i$ experimental points, including all species of targets $N$ and final-state hadrons $h$. 
In each $z$ bin for each replica the values of the collinear fragmentation functions $D_1^{a \smarrow h}$ are independently modified with a Gaussian noise with standard deviation equal to the theoretical error $\Delta D_1^{a\smarrow h}$. 
In this work we rely on different parametrizations for $D_1^{a \smarrow h}$: for pions we use the DSEHS analysis~\cite{deFlorian:2014xna} at NLO in $\alpha_S$; for kaons we use the DSS parametrization~\cite{deFlorian:2007aj} at LO in $\alpha_S$. 
The uncertainties $\Delta D_1^{a\smarrow h}$ are estimated from the plots in Ref.~\cite{Epele:2012vg}; they represents the only source of uncertainty in $\Delta m_{N,  {\rm theo}}^{h}$. 
Statistical and systematic experimental uncertainties $\Delta m_{N, \mbox{\tiny stat}}^{h}$ and $\Delta m_{N, \mbox{\tiny sys}}^{h}$ are taken from the experimental collaborations. 
We do not take into account the covariance among different kinematic bins. 

We minimize the error function in
Eq.~\eqref{e:MC_chi2} with \minuit~\cite{James:1975dr}. 
In each replica we randomize the starting point of the minimization, to better
sample the space of fit parameters.
 The final outcome is a set of $\mathcal{M}$ different vectors of best-fit parameters, $\{ p_{0r}\},\; r=1,\ldots \mathcal{M}$, with which we can calculate any observable, its mean, and its standard deviation. 
The distribution of these values needs not to be necessarily Gaussian. In fact, in this case the $1 \sigma$ confidence interval is different from the 68\% interval. 
The latter can simply be computed for each experimental point by rejecting the largest and the lowest 16\% of the $\mathcal{M}$ values.   

Although the minimization is performed on the function defined in
Eq.~\eqref{e:MC_chi2}, the agreement of the $\mathcal{M}$ replicas with the
original data is expressed in terms of a $\chi^2$ function defined as in
Eq.~\eqref{e:MC_chi2} but with the replacement $m_{N, r}^{h} \to m_{N}^{h}$,
i.e.,  with respect to the original data set. If the model is able to give a
good description of the data, the distribution of the $\mathcal{M}$ values of
$\chi^2$/d.o.f. should be peaked around one.  

\section{Results}
\label{s:results}

Our work aims at simultaneously fitting for the first time data sets related to different experiments. 
In the past, only fits related either to SIDIS or hadronic collisions have
been presented. Here we mention a selection of recent existing analyses. 

In Ref.~\cite{Signori:2013mda}, the authors fitted \hermes\ multiplicities
only (taking into account a total of 1538 points) 
without taking into account QCD evolution. 
In that work, a flavor decomposition in transverse momentum of the unpolarized
TMDs and an analysis of the kinematic dependence of the intrinsic average
square transverse momenta were presented.  
In Ref.~\cite{Anselmino:2013lza} the authors fitted \hermes\ and \compass\
multiplicities separately (576 and 6284 points respectively), without TMD
evolution and introducing an
ad-hoc normalization for \compass\ data. A fit of
SIDIS data including TMD evolution was performed on measurements by the H1
collaboration of the so-called transverse energy flow~\cite{Nadolsky:1999kb,Aid:1995we}. 
 
Looking at data from hadronic collisions, Konychev and
Nadolsky~\cite{Konychev:2005iy} fitted data of low-energy Drell--Yan events and
$Z$-boson production at Tevatron, taking into account TMD evolution at NLL
accuracy (this is the most recent of a series of important papers on
the subject~\cite{Ladinsky:1993zn,Landry:1999an,Landry:2002ix}). They fitted in total 98
points. Contrary to our approach, Konychev and Nadolsky studied the quality of the fit as a
function of $\bb_{\text{max}}$. They found that the best value for
$\bb_{\text{max}}$ is $1.5$ GeV$^{-1}$ (to be compared to our choice
$\bb_{\text{max}}\approx 1.123$ GeV$^{-1}$, see
Sec.~\ref{ss:TMDevo}). 
 Comparisons of best-fit values in the nonperturbative Sudakov form factors
 are delicate, since the functional form is different from ours. 
In 2014 D'Alesio, Echevarria, Melis, Scimemi performed a
fit~\cite{DAlesio:2014mrz} of Drell--Yan data and $Z$-boson production data at
Tevatron, focusing in particular on the role of the nonperturbative
contribution to the kernel of TMD evolution. This is the fit with the highest
accuracy in TMD evolution performed up to date (NNLL in the Sudakov exponent and ${\cal O}(\alpha_S)$ in the Wilson coefficients). 
In the same year Echevarria, Idilbi, Kang and Vitev~\cite{Echevarria:2014xaa}
presented a parametrization of the unpolarized TMD that described
qualitatively well some bins of \hermes\ and \compass\ data, together with
Drell--Yan and Z-production data.
A similar result was presented by Sun, Isaacson, Yuan and
Yuan~\cite{Su:2014wpa}.

In the following, we detail the results of a fit to the data sets described in
Sec.~\ref{s:data_analysis} with a flavor-independent configuration for the
transverse momentum dependence of unpolarized TMDs. 
In Tab.~\ref{t:fl_ind_chi2} we present the total $\chi^2$. The number of
degrees of freedom (d.o.f.) is given by the number of data points analyzed
reduced by the number of free parameters in the error function.  
The overall quality of the fit is good, with a global $\chi^2$/d.o.f. $= 1.55 \pm 0.05$. Uncertainties are computed as the $68\%$ confidence level (C.L.) from the replica methodology. 
\begin{table}[h!]
\small
  \centering
  \begin{tabular}{|c|c|c|c|}
\hline
\hline
Points& Parameters & $\chi^2$& $\chi^2/$d.o.f. \\
\hline
8059 & 11  & $12629 \pm 363$ & $1.55 \pm 0.05$ \\
\hline
\hline
\end{tabular}
\caption{Total number of points analyzed, number of free parameters and $\chi^2$ values.}
\label{t:fl_ind_chi2}
\end{table}

\subsection{Agreement between data and theory}
\label{ss:data_vs_theory}

The partition of the global $\chi^2$ among SIDIS off a proton, SIDIS off a
deuteron, Drell--Yan and $Z$ production events is given in
Tab.~\ref{t:fl_ind_chi2_eP},~\ref{t:fl_ind_chi2_eD},~\ref{t:fl_ind_chi2_DY},~\ref{t:fl_ind_chi2_Z}
respectively.

\subsubsection*{Semi-inclusive DIS}
\label{sss:SIDIS_agreement}

For SIDIS at \hermes\ off a proton, 
most of the contribution to the $\chi^2$ comes from events with a $\pi^+$ in the final state. 
In Ref.~\cite{Signori:2013mda} the high $\chi^2$ was attributed to the
poor agreement between experiment and theory at
the level of the collinear multiplicities. 
In this work we use a newer parametrization of the collinear FFs
(DSEHS~\cite{deFlorian:2014xna}), based on a fit which includes \hermes\
collinear pion multiplicities.  In spite of this improvement, the
contribution to $\chi^2$ from \hermes\ data is higher then in
Ref.~\cite{Signori:2013mda}, because the
present fit includes data from other experiments (\hermes\ represents less
than 20\% of the whole data set).
The bins with the worst agreement are at low $Q^2$. As we will discuss in
Sec.~\ref{ss:replica105}, we think that the main reason for the large $\chi^2$
at \hermes\ is a normalization difference. This may also be due to the fact
that we are computing our theoretical estimates at the average values of the
kinematic variables, instead of integrating the multiplicities in each bin. 
Kaon multiplicities have in general a lower $\chi^2$, due to the bigger
statistical errors and the large
uncertainties for the kaon FFs. 

\begin{table}[h!]
\begin{center}
\begin{tabular}{|c|c|c|c|c|}
 \hline
\hline
  & \hermes & \hermes & \hermes & \hermes \\   
 &  $p \to \pi^+$    &   $p \to \pi^-$    &  $p \to K^+$    &   $p \to K^-$
 \\
 \hline
 Points         &  190 & 190 & 189 & 187       \\
 \hline
$\chi^2 /$points & $4 .83\pm 0.42$ & $2 .47\pm 0.28$ & $0 .91\pm 0.14$ & $0 .82\pm 0.17$   \\            
\hline
\hline
\end{tabular}
\caption{Number of points analyzed and $\chi^2$ values for SIDIS off a proton target.}
\label{t:fl_ind_chi2_eP}
\end{center}
\end{table}

For pion production off a deuteron at \hermes\ the $\chi^2$ is lower with
respect to the production off a proton, but still compatible within
uncertainties. For kaon production off a deuteron the $\chi^2$ is higher with
respect to the scattering off a proton. The difference is especially large for $K^-$. 

SIDIS at \compass\ involves scattering off deuteron only, $D \to h^\pm$, and we identify $h \equiv \pi$. 
The quality of the agreement between theory and \compass\ data is better than
in the case of pion production at \hermes. This depends on at least two
factors: first, our fit is essentially driven by the \compass\ data, which
represent about 75\% of the whole data set; second, the
observable that we fit in this case is the normalized
multiplicity, defined in Eq.~\eqref{e:mult_norm}. This automatically eliminates
most of the discrepancy between theory and data due to normalization. 
\begin{table}[h!]
\begin{center}
\begin{tabular}{|c|c|c|c|c|c|c|}
 \hline
\hline
  & \hermes & \hermes & \hermes & \hermes & \compass & \compass\\
 ~          &  $D \to \pi^+$    &   $D \to \pi^-$    &  $D \to K^+$    &   $D \to K^-$      &  $D \to h^+$    &   $D \to h^-$            \\
\hline
 Points         &  190 & 190 & 189 & 189   & 3125 & 3127   \\
 \hline
$\chi^2 /$points & $3.46\pm 0.32$ & $2.00\pm 0.17$ & $1.31\pm 0.26$ & $2.54\pm 0.57$  & $1.11\pm 0.03$ & $1.61\pm 0.04$ \\            
 \hline
 \hline
\end{tabular}
\caption{Number of points analyzed and $\chi^2$ values for SIDIS off a deuteron target.} 
\label{t:fl_ind_chi2_eD}
\end{center}
\end{table}

\begin{figure}[h!]
\begin{center}
\includegraphics[width=0.90\textwidth]{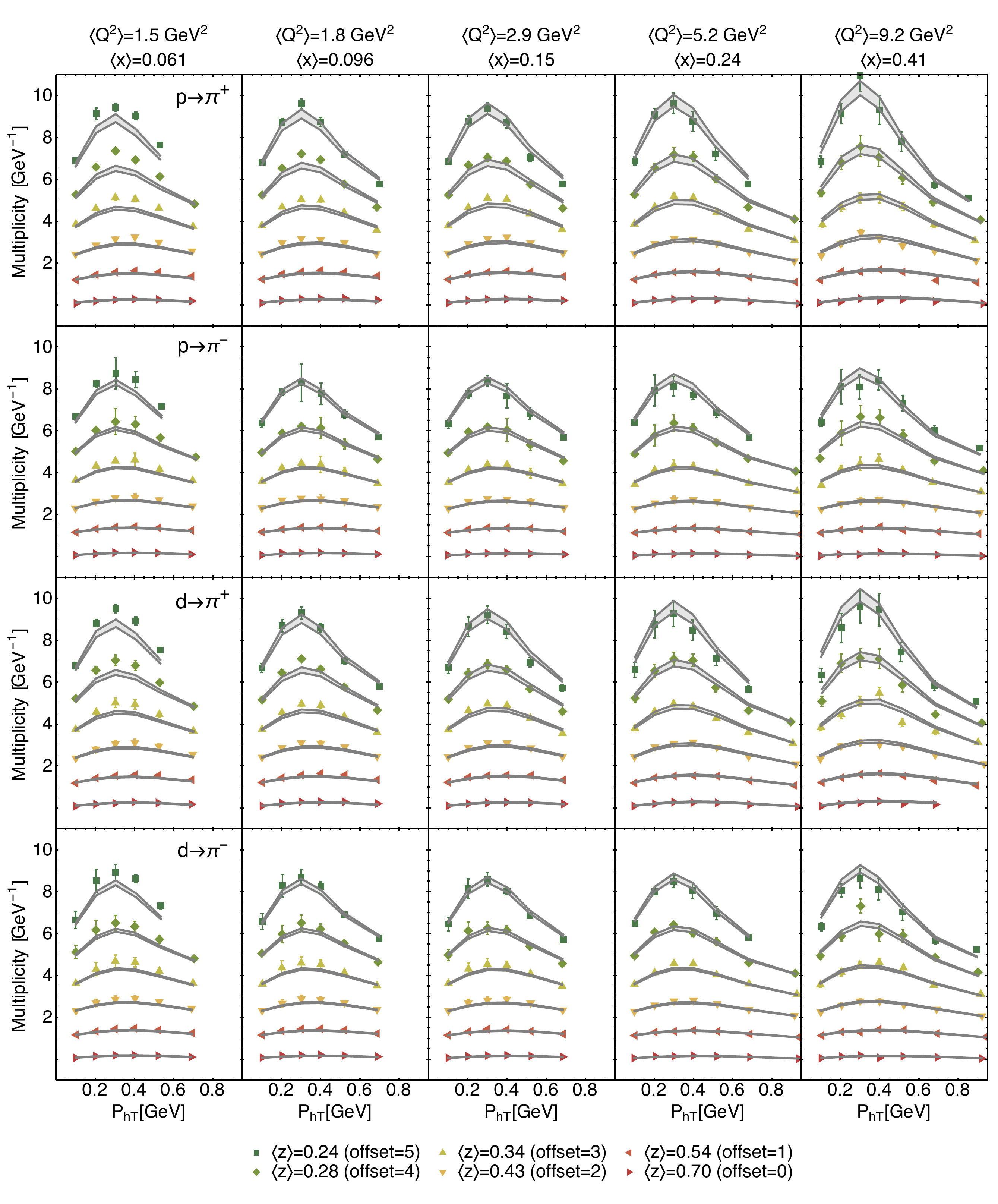}
\end{center}
\caption{\hermes\ multiplicities for production of pions off a proton and a deuteron for different $\langle x \rangle$, $\langle z \rangle$, and $\langle Q^2 \rangle$ bins as a function of the transverse momentum of the detected hadron  $P_{hT}$. For clarity, each $\langle z \rangle$  bin has been shifted by an offset indicated in the legend.} 
\label{f:H_pions}
\end{figure}

Fig.~\ref{f:H_pions} presents the agreement between the theoretical formula in~\eqref{e:multiplicity} and the \hermes\ multiplicities for production of pions off a proton and a deuteron. 
Different $\langle x \rangle$, $\langle z \rangle$ and $\langle Q^2 \rangle$
bins are displayed as a function 
of the transverse momentum of the detected hadron $P_{hT}$.
The grey bands are an envelope of the $200$ replica of best-fit curves. For every point in $P_{hT}$ we apply a $68\%$ C.L. selection criterion. 
Points marked with different symbols and colors correspond to different
$\langle z \rangle$ values. There is a strong correlation between $\langle x
\rangle$ and $\langle Q^2 \rangle$ that does not allow us to explore the $x$ and $Q^2$ dependence of the TMDs separately. 
Studying the contributions to the $\chi^2$/points as a function of the kinematics,  
we notice that the $\chi^2(Q^2)$ 
tends to improve as we move to higher $Q^2$ values, where the kinematic approximations of factorization are more reliable. 
Moreover, usually the $\chi^2(z)$ increases at lower $z$ values.

\begin{figure}[h!]
\begin{center}
\includegraphics[width=0.90\textwidth]{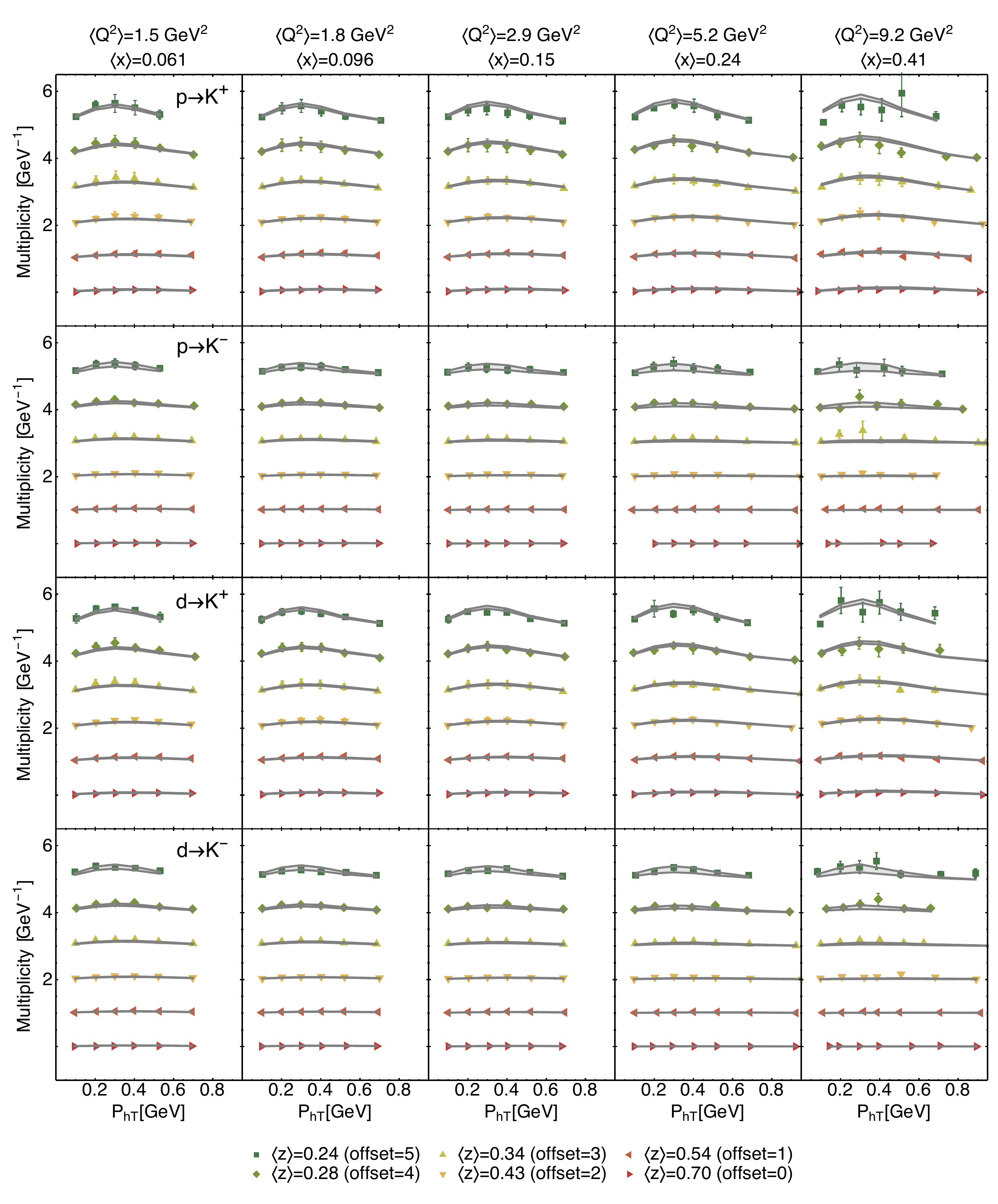}
\end{center}
\caption{\hermes\ multiplicities for production of kaons off a proton and a deuteron for different $\langle x \rangle$, $\langle z \rangle$, and $\langle Q^2 \rangle$ bins as a function of the transverse momentum of the detected hadron $P_{hT}$. For clarity, each $\langle z \rangle$  bin has been shifted by an offset indicated in the legend.} 
\label{f:H_kaons}
\end{figure}

Fig.~\ref{f:H_kaons} has same contents and notation as in Fig.~\ref{f:H_pions}
but for kaons in the final state. In this case, the trend of the agreement as
a function of $Q^2$ is not as clear as for the case of pions: good agreement
is found also at low $Q^2$.

\begin{figure}[h!]
\begin{center}
\includegraphics[width=\textwidth]{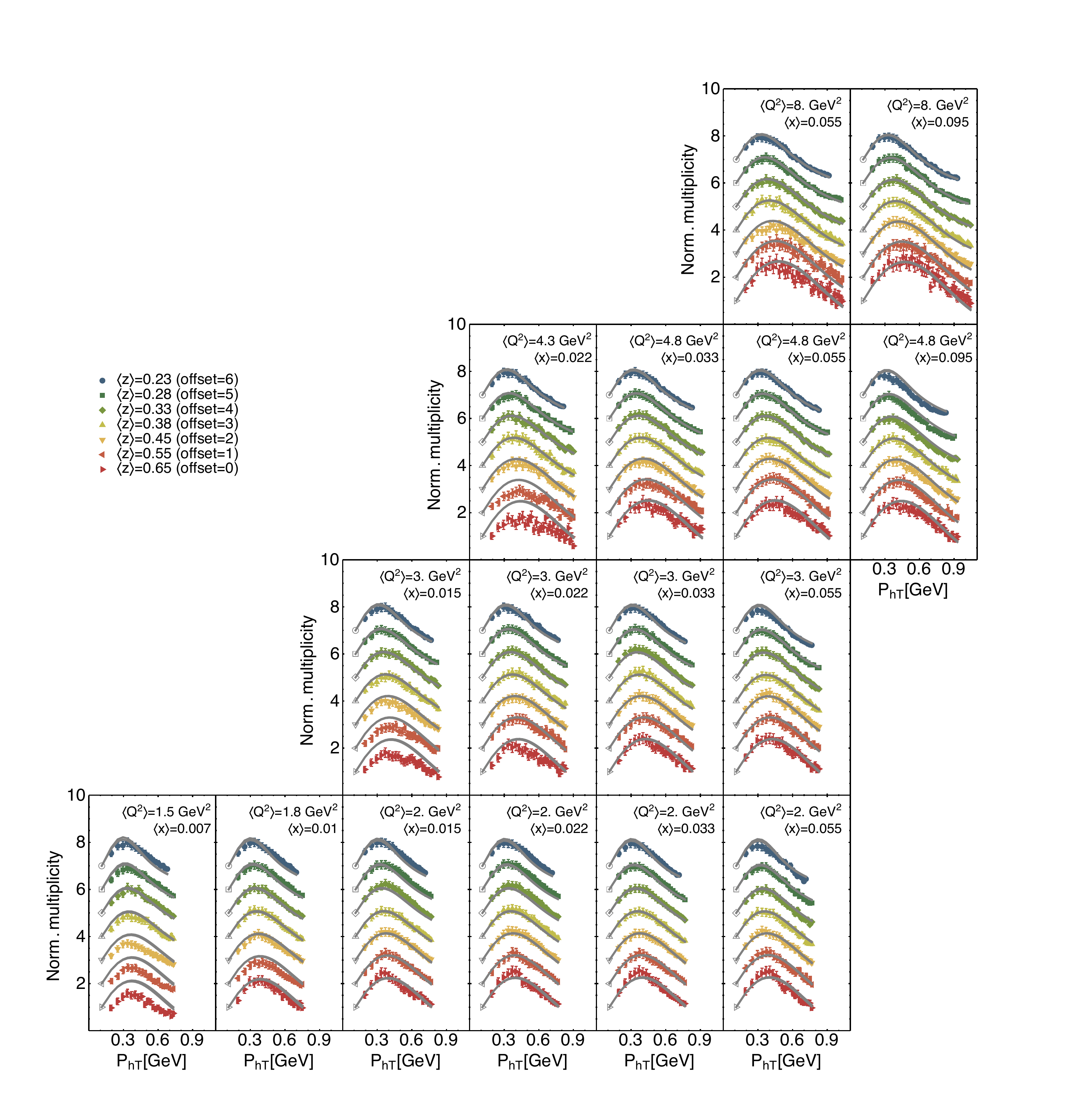}
\end{center}
\caption{\compass\ multiplicities for production of negative hadrons ($\pi^-$) off a deuteron for different $\langle x \rangle$, $\langle z \rangle$, and $\langle Q^2 \rangle$ bins as a function of the transverse momentum of the detected hadron  $P_{hT}$. Multiplicities are normalized to the first bin in $P_{hT}$ for each $\langle z \rangle$ value (see~\eqref{e:mult_norm}). For clarity, each $\langle z \rangle$  bin has been shifted by an offset indicated in the legend.} 
\label{f:C_pim}
\end{figure}

In Fig.~\ref{f:C_pim} we present \compass\ normalized multiplicities (see
Eq.~\eqref{e:mult_norm}) for production of $\pi^-$ off a deuteron for
different $\langle x \rangle$, $\langle z \rangle$, and $\langle Q^2 \rangle$
bins as a function of the transverse momentum of the detected hadron
$P_{hT}$. The open marker around the first $P_{hT}$ point in each panel indicates that the first value is fixed and not fitted. 
The correlation between $x$ and $Q^2$ is less strong than at \hermes\ and this
allows us to study different $\langle x \rangle$ bins at fixed $\langle Q^2 \rangle$.
For the highest $Q^2$ bins, the agreement is good for all $\langle x \rangle$, $\langle z \rangle$ and $P_{hT}^2$. 
In bins at lower $Q^2$, the descriptions gets worse, especially at low and
high $z$. 
For fixed $\langle Q^2 \rangle$ and high $\langle z \rangle$, a good agreement
is recovered moving to higher $\langle x \rangle$ bins (see, e.g., the
  third line from the top in Fig.~\ref{f:C_pim}).

\begin{figure}[h!]
\begin{center}
\includegraphics[width=\textwidth]{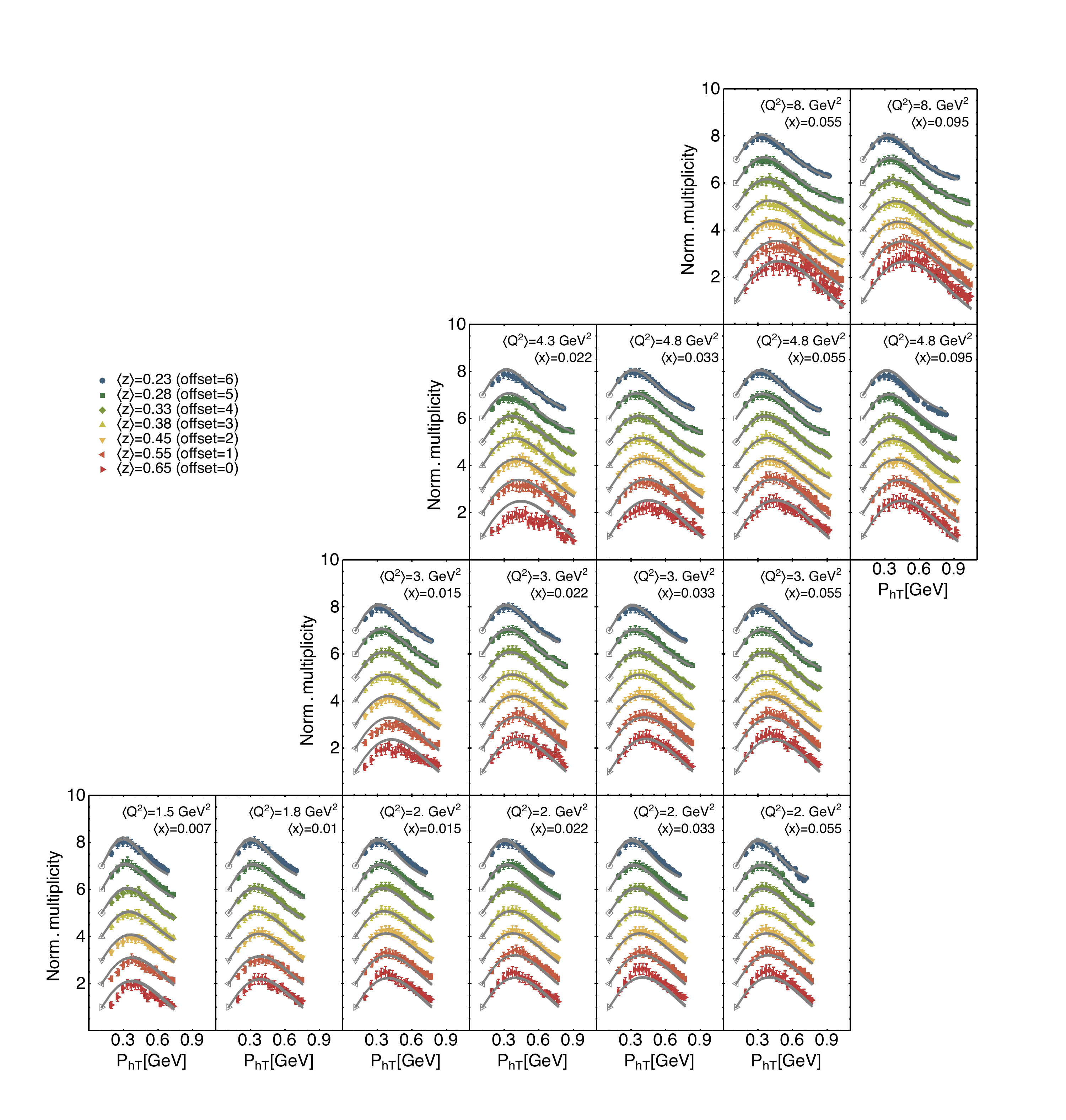}
\end{center}
\caption{\compass\ multiplicities for production of positive hadrons ($\pi^+$) off a deuteron for different $\langle x \rangle$, $\langle z \rangle$, and $\langle Q^2 \rangle$ bins as a function of the transverse momentum of the detected hadron $P_{hT}$. Multiplicities are normalized to the first bin in  $P_{hT}$ for each $\langle z \rangle$ value (see~\eqref{e:mult_norm}). For clarity, each $\langle z \rangle$  bin has been shifted by an offset indicated in the legend.} 
\label{f:C_pip}
\end{figure}

Fig.~\ref{f:C_pip} has same content and notation as in Fig.~\ref{f:C_pim}, but
for $h^+ \equiv \pi^+$. The same comments on the agreement between theory and
the data apply. 

\subsubsection*{Drell--Yan and $Z$ production}
\label{sss:DYZ_agreement}

\begin{figure}[h!]
\centering
\includegraphics[width=0.95\textwidth]{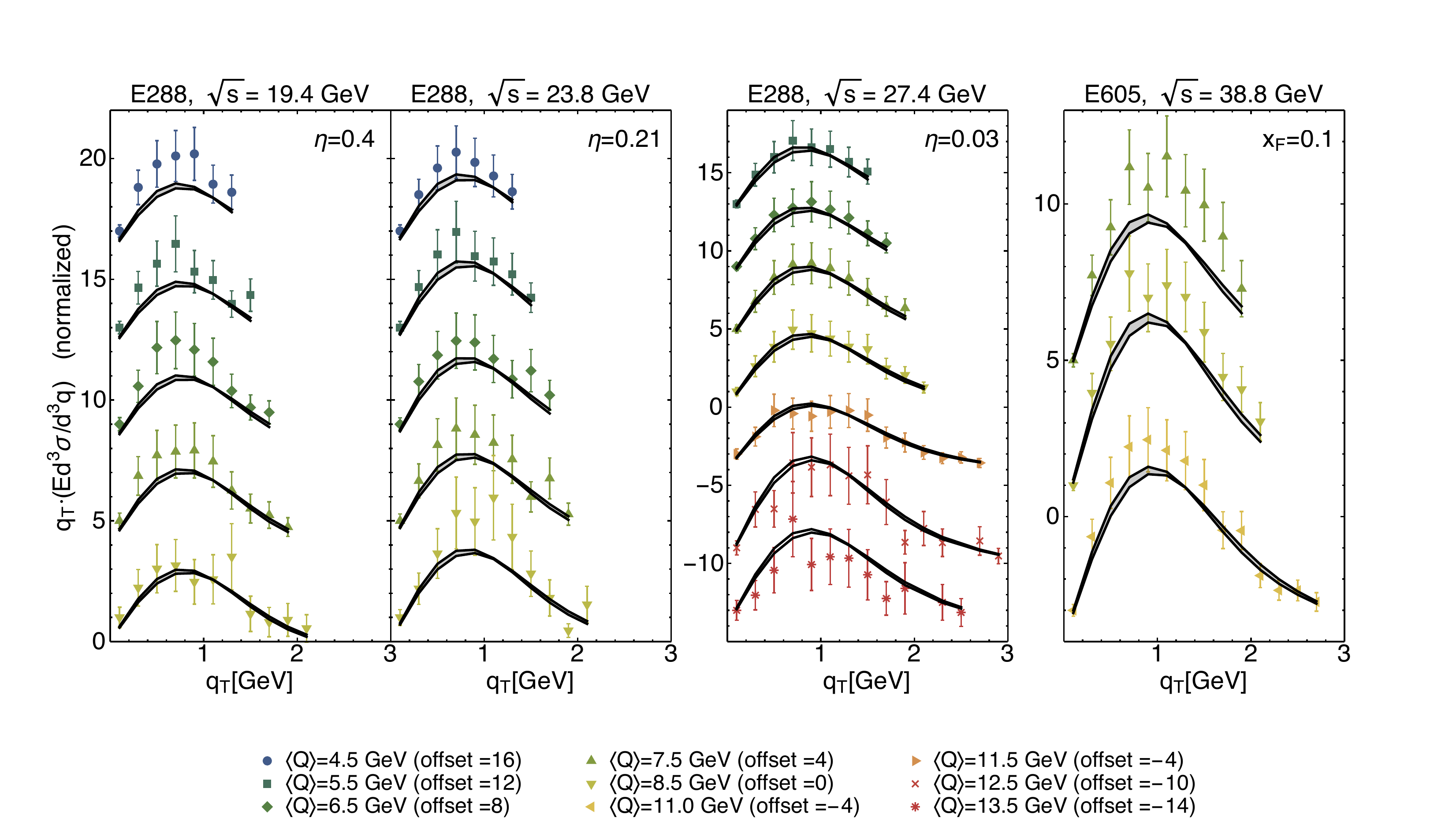}
\caption{Drell--Yan differential cross section for different experiments and
  different values of $\sqrt{s}$ and for different $\langle Q \rangle$
  bins. For clarity, each $\langle Q \rangle$ bin has been normalized 
  (the first data point has been set always equal to 1) and then shifted by an
  offset indicated in the legend.}
\label{f:DY_panel}
\end{figure}

The low energy Drell--Yan data collected by the E288 and E605 experiments at Fermilab have large error bands (see Fig.~\ref{f:DY_panel}). This is why the $\chi^2$ values in Tab.~\ref{t:fl_ind_chi2_DY} are rather low compared to the other data sets. 

The agreement is also good for $Z$ boson production, see Tab.~\ref{t:fl_ind_chi2_Z}. The statistics from Run-II is higher, which generates smaller experimental uncertainties and higher $\chi^2$, especially for the CDF experiment.
\begin{table}[h!]
\begin{center}
\renewcommand{\tabcolsep}{0.4pc} 
\renewcommand{\arraystretch}{1.2} 
\begin{tabular}{|c|c|c|c|c|}
 \hline
 \hline
 ~                        &  E288 [200]    &  E288 [300]        &  E288 [400]          &  E605                \\
 \hline
 Points                   &      45      &   45             &       78           &     35               \\
 \hline
$ \chi^2  /$points      &  $0.99\pm 0.09$        &    $0.84\pm 0.10$           &       $0.32\pm 0.01$ &   $1.12\pm 0.08$     \\
\hline
\hline
\end{tabular}
\caption{Number of points analyzed and $\chi^2$ values for fixed-target Drell--Yan experiments at low energy. The labels in square brackets were introduced in Sec.~\ref{ss:dy}.}
\label{t:fl_ind_chi2_DY}
\end{center}
\end{table}
\begin{table}[h!]
\begin{center}
\renewcommand{\tabcolsep}{0.4pc} 
\renewcommand{\arraystretch}{1.2} 
\begin{tabular}{|c|c|c|c|c|}
 \hline
\hline
 ~                        & CDF Run I    &  D0 Run I        & CDF Run II        & D0 Run II      \\
 \hline
 Points                   &      31      &   14             &       37          &        8       \\
 \hline
$\chi^2 /$points     &  $1.36\pm 0.00$        &    $1.11\pm 0.02$           &       $2.00\pm 0.02$         &   $1.73\pm 0.01$     \\
\hline
\hline
\end{tabular}
\caption{Number of points analyzed and $\chi^2$ values for $Z$ boson production at Tevatron.}
\label{t:fl_ind_chi2_Z}
\end{center}
\end{table}

Fig.~\ref{f:DY_panel} displays the cross section for DY events differential with respect to the transverse momentum $q_T$ of the virtual photon, its invariant mass $Q^2$ and rapidity $y$.  
As for the case of SIDIS, the grey bands are the $68\%$ C.L. envelope of the
200 replicas of the fit function. The four panels represents different values
for the rapidity $y$ or $x_F$ (see Eq.~\eqref{e:eta_xf}). In each panel, we have plots for different $Q^2$ values.
The lower is $Q$, the less points in $q_T$ we fit (see also Sec.~\ref{ss:dy}). 
The hard scale lies in the region $4.5 < \langle Q \rangle < 13.5$ GeV. This
region is of particular importance, since these ``moderate'' $Q$ values should
be high enough to safely apply factorization and, at the same time, low enough
in order for the nonperturbative effects to not be shaded by transverse
momentum resummation. 

\begin{figure}[h!]
\begin{center}
\includegraphics[width=0.95\textwidth]{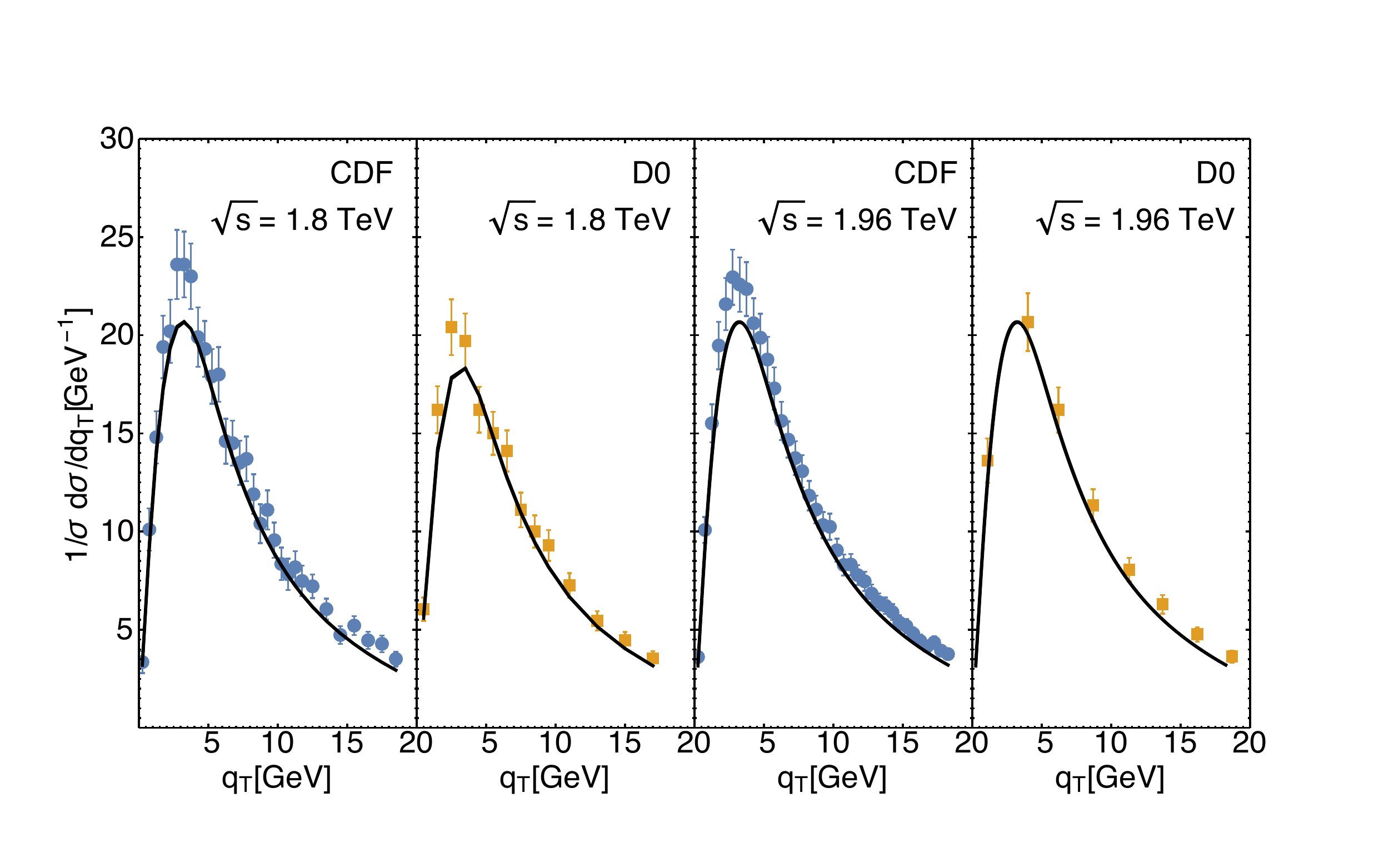}
\end{center}
\caption{Cross section differential with respect to the transverse momentum $q_T$ of a $Z$ boson produced from $p\bar{p}$ collisions at Tevatron. The four panels refer to different experiments (CDF and D$0$) with two different values for the center-of-mass energy ($\sqrt{s} = 1.8$ TeV and $\sqrt{s}=1.96$ TeV). In this case the band is narrow due to the narrow range for the best-fit values of $g_2$.} 
\label{f:Z_qT}
\end{figure}

In Fig.~\ref{f:Z_qT} we compare the cross section differential with respect to the transverse momentum $q_T$ of the virtual $Z$ (namely Eq.~\eqref{e:dsigma_gZ} integrated over $\eta$) with data from CDF and D0 at Tevatron Run I and II. 
Due to the higher $Q = M_Z$, the range explored in $q_T$ is much larger compared to all the other observables considered. The tails of the distributions deviate from a Gaussian behavior, as it is also evident in the bins at higher $Q^2$ in Fig.~\ref{f:DY_panel}. The band from the replica methodology in this case is much narrower, due to the reduced sensitivity to the intrinsic transverse momenta at $Q=M_Z$ and to the limited range of best-fit values for the parameter $g_2$, which controls soft-gluon emission. 
As an effect of TMD evolution, the peak shifts from $\sim 1$ GeV for Drell--Yan events in Fig.~\ref{f:DY_panel} to $\sim 5$ GeV in Fig.~\ref{f:Z_qT}. The position of the peak is affected both by the perturbative and the nonperturbative part of the Sudakov exponent (see Sec.~\ref{ss:TMDevo} and~\cite{Signori:2016lvd}).
Most of the contributions to the $\chi^2$ comes from normalization effects and
not from the shape in $q_T$ (see Sec.~\ref{ss:replica105}). 


\subsection{Transverse momentum dependence at 1 GeV}
\label{ss:bestfit_TMDs}

The variables $\bb_{\rm min}$ and $\bb_{\rm max}$ delimit the range in $\bT$
where transverse momentum resummation is computed perturbatively. 
The $g_2$ parameter enters the nonperturbative Sudakov exponent and
quantifies the amount of transverse momentum 
due to soft gluon radiation that is not included in the
perturbative part of the Sudakov form factor.
As already explained in Sec.~\ref{ss:TMDevo}, in this work we fix the value for
$\bb_{\rm min}$ and $\bb_{\rm max}$ in such a way that at $Q=1$ GeV the
unpolarized TMDs coincide with their nonperturbative input. We leave $g_2$ as 
a fit parameter. 

Tab.~\ref{t:fl_ind_parcommon} summarizes the chosen values of $\bb_{\rm min}$,
$\bb_{\rm max}$ and the best-fit value for $g_2$. The latter is given as an
average with $68\%$ C.L. uncertainty computed over the set of 200 replicas. We also quote the results obtained from
replica 105, since its parameters are very close to the
mean values of all replicas. 
We obtain a value $g_2=0.13 \pm 0.01$, smaller than the value ($g_2 = 0.184 \pm 0.018$) obtained in Ref.~\cite{Konychev:2005iy},
where however no SIDIS data was taken into consideration, and smaller than the
value ($g_2=0.16$) chosen in Ref.~\cite{Echevarria:2014xaa}. We stress however
that our prescriptions involving both $\bb_{\text{min}}$ and
$\bb_{\text{max}}$ are different from previous works.

Tab.~\ref{t:fl_ind_par_TMD} collects the best-fit values of parameters in the
nonperturbative part of the TMDs at $Q=1$ GeV (see Eqs.~\eqref{e:f1NP}
and~\eqref{e:D1NP}); as for $g_2$, we give the average value over the full set
of replicas and the standard deviation based on a $68\%$ C.L. (see
Sec.~\ref{ss:replica_method}), and we also quote the value of replica 105.

\begin{figure}[h!]
\begin{center}
\includegraphics[width=0.60\textwidth]{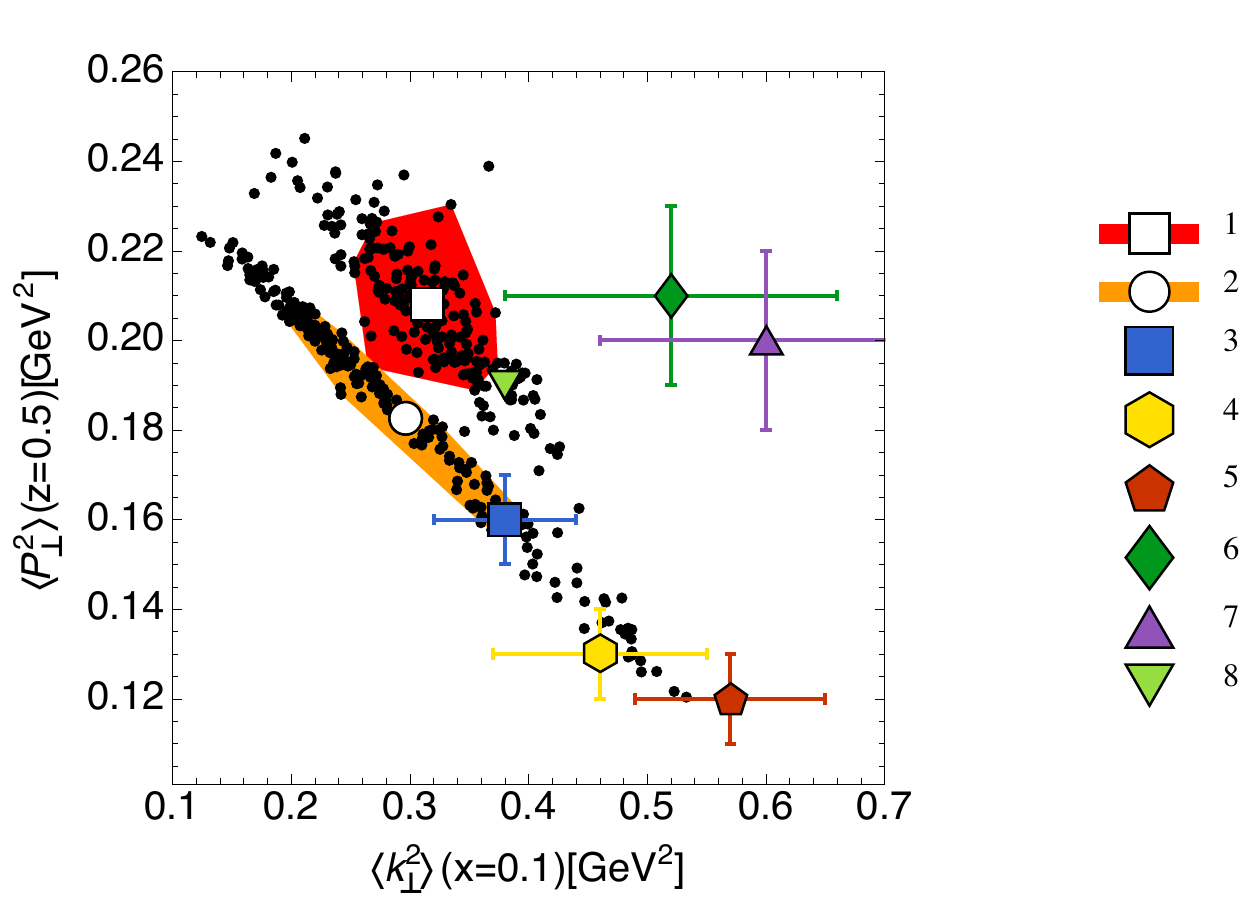}
\end{center}
\caption{Correlation between transverse momenta in TMD FFs, $\langle P_\perp^2
  \rangle(z=0.5)$, and in TMD PDFs, $\langle k_\perp^2 \rangle(x=0.1)$, in
  different phenomenological extractions. 
 (1): average values (white square) obtained in the present analysis, values
 obtained from each replica (black dots) and
 $68\%$ C.L. area (red); (2) results from Ref.~\cite{Signori:2013mda},
 (3) results from Ref.~\cite{Schweitzer:2010tt}, (4) results from Ref.~\cite{Anselmino:2013lza} for
 \hermes\ data, 
 (5) results from Ref.~\cite{Anselmino:2013lza} for \hermes\ data at high $z$, (6) results from Ref.~\cite{Anselmino:2013lza} for normalized \compass\ data, (7) results from Ref.~\cite{Anselmino:2013lza} for normalized \compass\ data at high $z$, (8) results from Ref.~\cite{Echevarria:2014xaa}.} 
\label{f:kT2_vs_PT2}
\end{figure}

\begin{figure}[h!]
\centering
\begin{tabular}{ccc}
\includegraphics[width=0.40\textwidth]{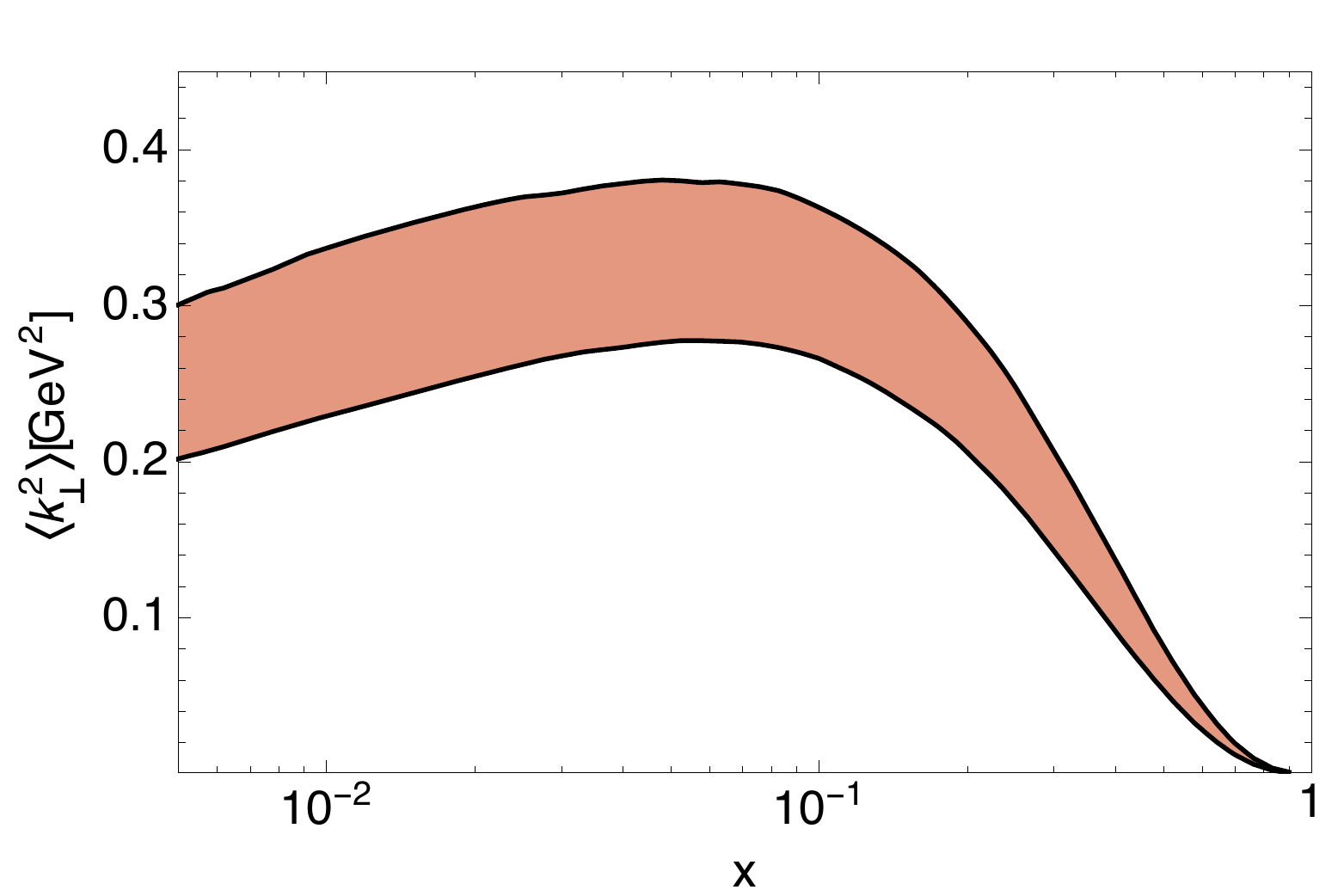}
&\hspace{0.001cm}
&
\includegraphics[width=0.40\textwidth]{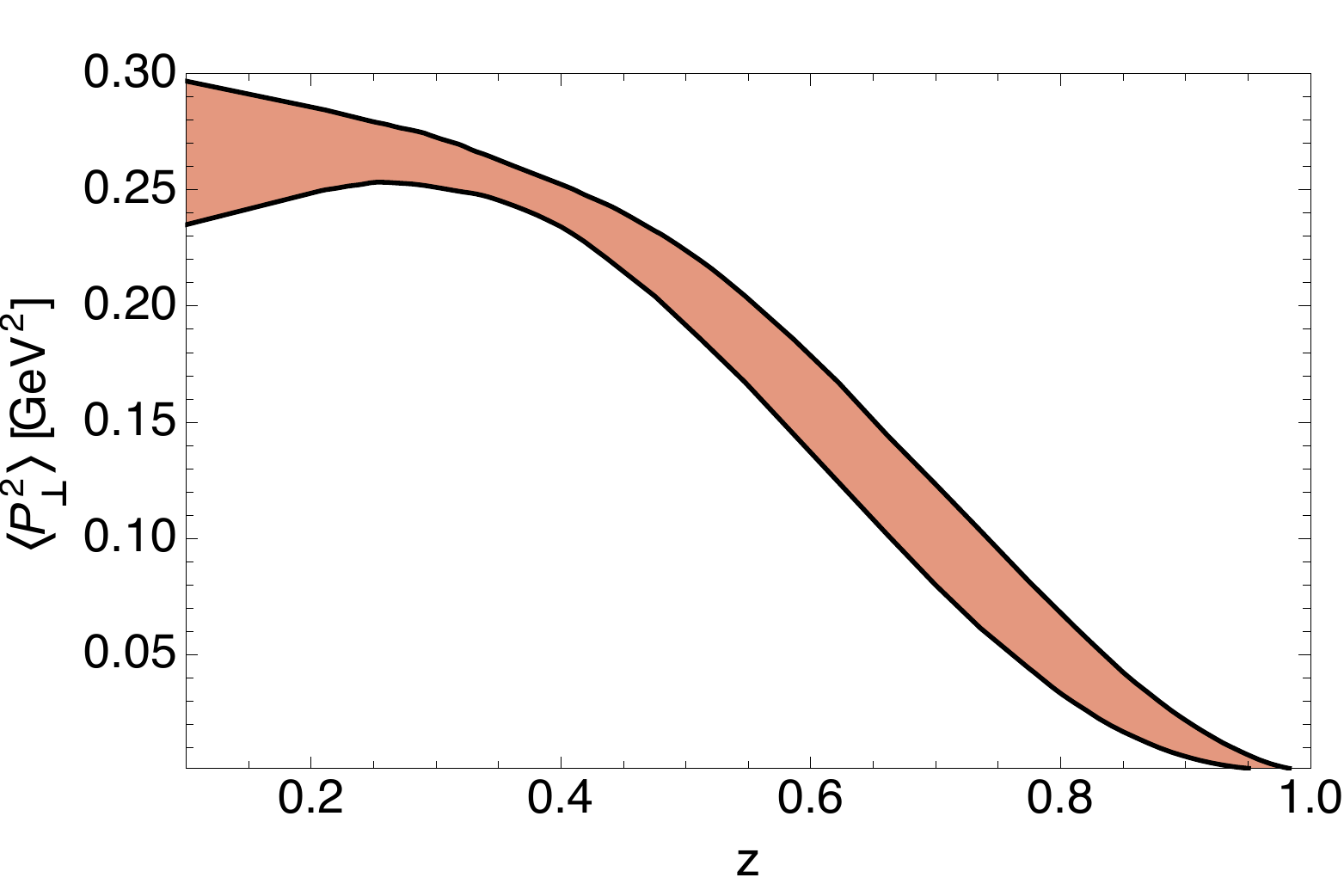}
\\
(a) && (b)
\end{tabular}
\caption{Kinematic dependence of $\big \langle \bm{k}_{\T}^2 \big \rangle (x)$
  (a) and of $\big \langle \bm{P}_{\perp}^2 \big \rangle (z)$ (b). The bands
  are the $68\%$ C.L. envelope of the full sets of best-fit
  curves.  
}
\label{f:avmomenta_68CL}
\end{figure}

In Fig.~\ref{f:kT2_vs_PT2} we compare different extractions of
partonic transverse momenta. The horizontal axis shows 
the value of the average transverse
momentum squared for the incoming parton, $\big \langle \bm{k}_{\T}^2 \big \rangle (x=0.1)$ (see Eq.~\eqref{e:transmom2}). 
The vertical axis shows the value of $\big \langle \bm{P}_{\perp}^2 \big \rangle (z=0.5)$, 
the average transverse momentum squared acquired during the fragmentation process (see Eq.~\eqref{e:transmom2}).  
The white square (label 1) indicates the average values of the two quantities
obtained in the present analysis at $Q^2=1$ GeV$^2$.
Each black dot around the white square is an outcome of one replica. The red
region around the white square contains the $68\%$ of the replicas that are closest to the average value.
The same applies to the white circle and the orange region around it (label 2),
related to the flavor-independent version of the analysis in
Ref.~\cite{Signori:2013mda}, obtained by fitting only \hermes\ SIDIS
data at an average $\langle Q^2 \rangle= 2.4$ GeV$^2$ and neglecting QCD evolution. 
A strong anticorrelation between the transverse momenta is evident in this
older analysis. 
In our new analysis, the inclusion of Drell--Yan and $Z$ production data adds physical information
about TMD PDFs, free from the influence of TMD FFs. This reduces significantly the 
correlation between $\big \langle \bm{k}_{\T}^2 \big \rangle (x=0.1)$ and $\big \langle \bm{P}_{\perp}^2 \big \rangle (z=0.5)$.  
The $68\%$ confidence region is smaller than in the older analysis. 
The average values of $\big \langle \bm{k}_{\T}^2 \big \rangle (x=0.1)$ are similar and compatible within error bands. 
The values of $\big \langle \bm{P}_{\perp}^2 \big \rangle (z=0.5)$ in the present analysis
turn out to be larger than in the older analysis, an effect that is due mainly to \compass\ data.
It must be kept in mind that the two analyses lead also to differences in the $x$ and
$z$ dependence of the transverse momentum squared. 
This dependence is shown in Fig.~\ref{f:avmomenta_68CL} (a) for $\big
\langle \bm{k}_{\T}^2 \big \rangle (x)$ and Fig.~\ref{f:avmomenta_68CL} (b)
for $\big \langle \bm{P}_{\perp}^2 \big \rangle (z)$. 
The bands are computed as the $68\%$ C.L.  envelope of the full sets of curves from the 200 replicas. Comparison with other extractions are presented and the legend is detailed in the caption of Fig.~\ref{f:kT2_vs_PT2}.

\begin{table}[h!]
\small
  \centering
  \begin{tabular}{|c|c|c|c|}
\hline
\hline
&$\bb_{\rm max}$ [GeV$^{-1}$] & $\bb_{\rm min}$ [GeV$^{-1}$] &  $g_2$ {[GeV$^2$]} 
 \\ 
& (fixed)     & (fixed)   &                            \\
\hline
All replicas & $2 e^{-\gamma_E}$& $2 e^{-\gamma_E}/Q$  & $0.13 \pm 0.01$  \\
\hline
Replica 105 &  $2 e^{-\gamma_E}$& $2 e^{-\gamma_E}/Q$  & $0.128$  \\
\hline
\hline
\end{tabular}
\caption{Values of parameters common to TMD PDFs and TMD FFs.}
\label{t:fl_ind_parcommon}
\end{table}
\begin{table}[h!]
\small
  \centering
  \begin{tabular}{|c||c|c|c|c|c|c|}
\hline
\hline
TMD PDFs&  $g_1$ 
& $\alpha$ & $\sigma$ & & $\lambda$ &  
 \\ 
        & {[GeV$^2$]}                               &
       &      &  &{[GeV$^{-2}$]} & \\
\hline
All replicas &  $0.28\pm 0.06$ & $2.95\pm 0.05$ & $0.17\pm 0.02$ & 
                & $0.86\pm 0.78$ & 
\\
\hline
Replica 105  &  $0.285$ & $2.98$ & $0.173$ & & $0.39$ & \\
\hline
\hline
TMD FFs&  $g_3$ &
$\beta$ & $\delta$ & $\gamma$ & $\lambda_F$ & $g_4$
 \\ 
        & {[GeV$^2$]} &            &        & &{[GeV$^{-2}$]} &{[GeV$^2$]}    \\
\hline
All replicas & $0.21\pm 0.02$ & $1.65\pm 0.49$ & $2.28\pm 0.46$ & $0.14\pm 0.07$ &
$5.50\pm 1.23$ & $0.13\pm 0.01$ \\
\hline
Replica 105   &  
 $0.212$ & $2.10$ & $2.52$ & $0.094$ & $5.29$ & $0.135$ \\
\hline
\hline
\end{tabular}
\caption{68\% confidence intervals of best-fit values for parametrizations of TMDs at $Q=1$ GeV.}
\label{t:fl_ind_par_TMD}
\end{table}

\subsection{Stability of our results}
\label{ss:replica105}

In this subsection we discuss the effect of modifying some of the choices we
made in our default fit. Instead of repeating the fitting procedure with
different choices, we limit ourselves to checking how the $\chi^2$ of a single
replica is affected by the modifications. 

As starting point we choose replica
105, which, as discussed above, is one of the most representative among the
whole replica set. 
The global $\chi^2/$d.o.f.\ of replica 105 is 1.51. We keep all parameters
fixed, without performing any new minimization, 
and we compute the $\chi^2/$d.o.f.\ after the modifications described in the
following.

First of all, we analyze \hermes\ data with the same strategy as \compass,
i.e., we normalize \hermes\ data to the value of the first bin
in $P_{hT}$. In this case, the global
$\chi^2/$d.o.f.\ reduces sharply to 1.27. The partial $\chi^2$ for the
different SIDIS processes measured 
at \hermes\ are shown in Table~\ref{t:replica105-hermes}. 
This confirms that normalization effects are the main contribution to the
$\chi^2$ of SIDIS data and have minor effects on TMD-related parameters. In
fact, even if we perform a new fit with this modification, the $\chi^2$ does
not improve significantly and parameters do not change much.

\begin{table}[h!]
\begin{center}
\begin{tabular}{|c|c|c|c|c|c|c|c|c|}
 \hline
\hline
 ~     &  $p \to \pi^+$    &   $p \to \pi^-$    &  $p \to K^+$    &   $p \to K^-$       &  $D \to \pi^+$    &   $D \to \pi^-$    &  $D \to K^+$    &   $D \to K^-$                \\
\hline
 Original   &  5.18 &  2.67 & 0.75  & 0.78      &  3.63 &  2.31 & 1.12  & 2.27    \\
 \hline
Normalized  &  1.94 &  1.13 &  0.57 & 0.29 & 1.59  & 0.80 & 0.47 & 0.97  \\            
 \hline
 \hline
\end{tabular}
\caption{$\chi^2$/d.o.f.\ for \hermes\ data with and without normalization to the value of the first bin in $P_{hT}$.} 
\label{t:replica105-hermes}
\end{center}
\end{table}

We consider the effect of changing the normalization of the $Z$-boson
data: if we increase the normalization factors quoted
in the last row of Tab.~\ref{t:data_Z} by 5\%, the $\chi^2$ quoted in the last
row of Tab.~\ref{t:fl_ind_chi2_Z}
drops to 0.66, 
0.52, 0.65, 0.68. This effect is
also already visible by eye in Fig.~\ref{f:Z_qT}: the theoretical curves are
systematically below the experimental data points, but the shape is reproduced
very well.

We consider the sensitivity of our results to the 
 parameterizations adopted for the collinear quark PDFs. 
The $\chi^2/ \text{d.o.f.}$ varies from its original value
1.51, obtained with the NLO GJR 2008 parametrization~\cite{Gluck:2007ck}, to
1.84 using NLO MSTW 2008~\cite{Martin:2009iq}, and 1.85 using NLO
CJ12~\cite{Owens:2012bv}. In both cases, the agreement with \hermes\
and $Z$ boson data is not affected significanlty,
 the agreement with \compass\ data becomes slightly worse, 
and the agreement with DY data becomes clearly worse. 

An extremely important point is the choice of kinematic cuts. Our default
choices are listed in Tabs.~\ref{t:data_SIDIS_proton}--\ref{t:data_Z}. We
consider also more stringent kinematic cuts on SIDIS data: 
$Q^2 > 1.5$ GeV$^2$ and 
$0.25 < z < 0.6$ instead of $Q^2 > 1.4$ GeV$^2$ and $0.2 < z <
0.7$, leaving the other ones unchanged. The number of bins with these cuts 
reduces from 8059 to 5679 and 
the  $\chi^2/ \text{d.o.f.}$  decreases to the
value 1.23. 
In addition, if we replace the constraint  $P_{h T} < \text{Min} [
0.2\, Q, 0.7\, Q z] + 0.5$ GeV  
with $P_{h T} < \text{Min} [ 0.2\, Q, 0.5\, Q z] +
0.3$ GeV, the number of bins reduces to 3380 and the $\chi^2/$d.o.f.\ decreases
further to 0.96. By adopting the even stricter 
cut $P_{h T} < 0.2\, Q z$, 
the number of bins drops to only 477, with  a 
 $\chi^2$/d.o.f.\ =1.02.  We can conclude that our fit, obtained by
 fitting data in an extended kinematic region, where TMD factorization may be
 questioned,  works extremely well also in a narrower
 region, where TMD factorization is expected to be under control.

\section{Conclusions}
\label{s:conclusions}

In this work we demonstrated for the first time that it is possible 
to perform a
simultaneous fit of unpolarized TMD PDFs and FFs 
to data of SIDIS, Drell--Yan and $Z$
boson production at small transverse momentum collected by different experiments. 
This constitutes the first attempt towards a global fit of 
$f_1^a(x,k_\perp^2)$ and $D_1^{a \to h}(z,P_\perp^2)$ in the context of TMD
factorization and with the implementation of 
TMD evolution at NLL accuracy. 
  
We extracted unpolarized TMDs using 8059 data points with 11 free parameters
using a replica methodology. We selected data with 
$Q^2 > 1.4$ GeV$^2$ and $0.2 < z < 0.7$. We restricted our fit to the small
transverse momentum region, selecting the maximum value of transverse momentum
on the basis of phenomenological considerations
(see Sec.~\ref{s:data_analysis}). With these choices, 
we included regions where TMD
factorization could be questioned, but we checked that our results describe 
very well the regions where TMD factorization is supposed to hold. 
The average $\chi^2$/d.o.f. is $1.55 \pm 0.05$ and can be improved up to 1.02
restricting the kinematic cuts, without changing the parameters (see Sec.~\ref{ss:replica105}). 
Most of the discrepancies between experimental data and theory comes from the
normalization and not from the transverse momentum shape. 

Our fit is performed assuming that the intrinsic transverse momentum
dependence of TMD PDFs and FFs 
can be parametrized by a normalized linear combination of a Gaussian and a
weighted Gaussian. We considered that the widths of the Gaussians depend on the longitudinal
momenta. We neglected a possible flavor dependence. 
For the nonperturbative component of TMD evolution, we adopted the choice most
often used in the literature (see Sec.~\ref{ss:TMDevo}). 

We plan to release grids of the parametrizations studied in this work via
TMDlib~\cite{Hautmann:2014kza} to facilitate phenomenological studies for
present and future experiments. 

In future studies, different
functional forms for all the nonperturbative ingredients should be explored, 
including also a possible
flavor dependence of the intrinsic transverse momenta.
A more
precise analysis from the perturbative point of view is also needed, which
should in principle make it possible to relax the tension in the normalization and to describe data at higher transverse
momenta. Moreover, the description at low transverse momentum  
should be properly matched to 
the collinear fixed-order calculations at high transverse momentum. 

Together with an improved theoretical framework, in order to better understand
the formalism more experimental data is needed. It would be particularly
useful to extend the
coverage in $x$, $z$, rapidity, and $Q^2$.  
The $12$ GeV physics program at Jefferson Lab~\cite{Dudek:2012vr} will be very important to constrain TMD distributions at large $x$.
Additional data from SIDIS (at \compass, at a future Electron-Ion
Collider), Drell--Yan (at \compass, at Fermilab), 
$Z/W$ production (at LHC, RHIC, and at
A Fixed-Target Experiment at the LHC~\cite{Brodsky:2012vg}) 
will be very important. Measurements
related to unpolarized
TMD FFs at $e^+e^-$ colliders (at Belle-II, BES-III, at a
future International Linear Collider) will be invaluable, since they are
presently missing.

Our work focused on quark TMDs.
We remark that at present almost nothing is known experimentally about gluon
TMDs~\cite{Mulders:2000sh,Echevarria:2015uaa}, because they typically require
higher-energy scattering processes and they are harder to isolate as compared
to quark distributions. Several promising measurements have been proposed in
order to extract both the unpolarized and linearly polarized gluon TMDs inside
an unpolarized proton. The cleanest possibility would be to look at dijet and
heavy quark pair production in electron-proton collisions at a future
EIC~\cite{Boer:2010zf,Pisano:2013cya}. Other proposals include isolated
photon-pair production at RHIC~\cite{Qiu:2011ai} and quarkonium production at
the LHC~\cite{Boer:2012bt,Dunnen:2014eta,Signori:2016jwo,Lansberg:2017tlc}.   

Testing the formalism of TMD factorization and understanding the structure of unpolarized TMDs is only the first crucial step in the exploration of the 3D proton structure in momentum space and this work opens the way to global determinations of TMDs. 
Building on this, we can proceed to deepen our understanding of hadron
structure via polarized structure function and asymmetries (see, e.g.,
Refs.~\cite{Aschenauer:2015ndk,Boglione:2015zyc} and references therein) and, at the
same time, to test the impact of hadron structure in precision measurements at
high-energies, such as at the LHC. A detailed mapping of hadron structure is
essential to interpret data from hadronic collisions, which are among the most
powerful tools to look for footprints of new physics.

\begin{acknowledgments}
Discussions with Giuseppe Bozzi are gratefully acknowledged.
This work is supported by the European Research Council (ERC) under the European Union's Horizon 2020 research and innovation program (grant agreement No. 647981, 3DSPIN). 
AS acknowledges support from U.S. Department of Energy contract DE-AC05-06OR23177, under which Jefferson Science Associates, LLC, manages and operates Jefferson Lab. 
The work of AS has been funded partly also by the program of the Stichting voor Fundamenteel Onderzoek der Materie (FOM), which is financially supported by the Nederlandse Organisatie voor Wetenschappelijk Onderzoek (NWO).
\end{acknowledgments}
\bibliographystyle{apsrevM}
\bibliography{mybiblio}

\end{document}